\documentclass[letterpaper,twocolumn,10pt]{article}
\usepackage{usenix}

\usepackage{tikz}
\usetikzlibrary{shapes}
\usetikzlibrary{positioning}
\usepackage{balance}
\usepackage{graphicx}%
\usepackage{amsmath}
\usepackage{subcaption}
\usepackage[export]{adjustbox}
\usepackage{booktabs}
\usepackage{tikz}
\usepackage{amsmath}
\usepackage{cleveref}
\usepackage{url}
\usepackage{caption}
\usepackage[most]{tcolorbox}
\usepackage{enumitem}
\usepackage{url}

\newcommand\Paragraph[1]{\smallskip \noindent\textbf{#1}.}
\begin{document}

\date{}

\title{Deanonymizing Ethereum Validators: The P2P Network Has a Privacy Issue}
\author{
{\rm Lioba Heimbach$^*$}\\
ETH Zurich \\
hlioba@ethz.ch
\and
{\rm Yann Vonlanthen$^*$}\\
ETH Zurich \\
yvonlanthen@ethz.ch
\and
{\rm Juan Villacis}\\
University of Bern\\
juan.villacis@unibe.ch
\and
{\rm Lucianna Kiffer}\\
IMDEA Networks\\
lucianna.kiffer@imdea.org
\and
{\rm Roger Wattenhofer}\\
ETH Zurich \\
wattenhofer@ethz.ch
} %

\maketitle
\def\thefootnote{*}\footnotetext{These authors contributed equally to this work.}
\def\thefootnote{\arabic{footnote}}

\begin{abstract}
Many blockchain networks aim to preserve the anonymity of validators in the \textit{peer-to-peer (P2P)} network, ensuring that no adversary can link a validator's identifier to the IP address of a peer due to associated privacy and security concerns.
This work demonstrates that the Ethereum P2P network does not offer this anonymity. We present a methodology that enables any node in the network to identify validators hosted on connected peers and empirically verify the feasibility of our proposed method. Using data collected from four nodes over three days, we locate more than 15\% of Ethereum validators in the P2P network. The insights gained from our deanonymization technique provide valuable information on the distribution of validators across peers, their geographic locations, and hosting organizations. We further discuss the implications and risks associated with the lack of anonymity in the P2P network and propose methods to help validators protect their privacy.
The Ethereum Foundation has awarded us a bug bounty, acknowledging the impact of our results.
\end{abstract}

\section{Introduction}

Ethereum is a blockchain that emphasizes decentralization, aiming to keep its consensus mechanism accessible to many participants, which contributes significantly to the complexity of its protocol. In particular, Ethereum faces challenges in scaling its consensus protocol while remaining accessible to smaller participants. The large number of validators involved in the consensus process and their extensive message exchanges lead to unprecedented complications. To address this challenge, innovative scaling solutions for the \textit{peer-to-peer (P2P)} network have been proposed and implemented~\cite{vyzovitis2022gossipsub}.

Our work demonstrates the impact of these scaling solutions on the privacy and security of the Ethereum P2P network and blockchain. We outline how to deanonymize validators in the P2P network by mapping a validator's identifier to the IP address of the machine it is hosted on. Our technique relies solely on observing \textit{attestation} (i.e., consensus layer vote) messages received from peers (i.e., nodes with established TCP connections). By analyzing messages from a peer $p$, we can infer whether a validator $v$ is hosted on this peer. %

Concretely, the main vulnerability stems from the current broadcast implementation, in which nodes are only responsible for propagating a pre-determined subset of all attestations. Thus, when a peer $p$ sends an attestation created by validator $v$ that falls outside their broadcasting responsibility, we can infer that the attestation was produced by $p$ itself. If we observe this behavior repeatedly, we demonstrate that with high confidence the attesting validator $v$ is connected to the peer $p$.

The Ethereum P2P network's privacy issue poses a major security risk, allowing attackers to identify nodes associated with validators set to create new blocks. This could lead to (D)DoS attacks, halting chain progress, or more targeted attacks on nodes associated with validators handling high value blocks, letting a subsequent malicious validator scoop these profits. We hope our work highlights this lack of privacy and informs future privacy-enhancing solutions.

\Paragraph{Contributions} We summarize our main contributions: 
\begin{itemize}[leftmargin=*,topsep=0pt,itemsep=-1ex,partopsep=1ex,parsep=1ex]
    \item We propose a simple and low-cost technique for a node in the network to deanonymize its peers, i.e., infer which validators they host. 
    \item We perform a measurement study to demonstrate the feasibility of the deanonymization. Using four nodes in just three days, we can locate more than 15\% of validators in the P2P network. 
    \item We outline the implication of the lack of anonymity in Ethereum's P2P network (e.g., fairness, liveness and safety concerns) and discuss possible mitigations. 
    \item Finally, we expose novel security risks in the P2P network, highlighting how validators concentrate on certain peers (e.g., we locate over 19,000 validators on a single peer) and how they are spread globally and across organizations (i.e., cloud service and internet service providers). We also discover that operators for different staking pools run multiple pools’ validators on the same machine, creating undesirable dependencies.\end{itemize}

\begin{tcolorbox}[title=Responsible Disclosure, colback=gray!5!white,colframe=gray!75!gray,boxsep=2pt,left=2pt,right=2pt,top=2pt,bottom=2pt]

We submitted a bug report to the Ethereum Foundation, disclosing the vulnerabilities presented in this paper. The Ethereum Foundation awarded us a bug bounty, acknowledging the impact of our results. We discuss further ethical considerations in \Cref{sec:ethics}.

\end{tcolorbox}

\section{Background}\label{sec:background}

In the following, we provide an overview of the Ethereum blockchain and its P2P network. 

\subsection{The Ethereum Blockchain}

The Ethereum blockchain operates as a Proof-of-Stake blockchain, and consists of the consensus layer (also called the Beacon chain) and the execution layer. Individuals can pay 32~ETH to become a \textit{validator}, i.e., a participant in the consensus algorithm. The 32~ETH represent a validator's \textit{stake} and can be confiscated in case of inactivity or malicious behavior. 
An individual can either run their own validator(s), i.e., \textit{solo/home stake} or stake through a \textit{staking pool}. Staking pools allow participants to pool resources to partake in the consensus algorithm with arbitrarily sized amounts of stake.\footnote{Note that the staking pool is responsible for fully operating the validator(s), unlike in Proof-of-Work mining pools.}

The \textit{consensus algorithm} is divided into \textit{epochs}, with each epoch lasting 32 \textit{slots} \cite{ethereum2024consensusspecs}. A slot lasts 12 seconds, resulting in an epoch duration of 6 minutes and 24 seconds. A slot corresponds to one new block appended to the blockchain. Among other validity rules, each block must be created (or \textit{proposed}) by the corresponding validator assigned to it and have sufficient \textit{attestations} of its correctness from the group (or \textit{commitee)} of validators assigned to it. We give an overview of relevant details below.

\Paragraph{Validators} Three main duties fall on validators~\cite{ethereum2024danebspecs}. First, during each \textit{slot} one validator is selected to propose a new block. Second, during each \textit{epoch}, each validator is assigned to a committee in one slot to make an attestation. Finally, with some probability, a validator is chosen as an aggregator for a slot they are attesting, i.e., they are tasked with collecting attestations and publishing an aggregate. Validators are identified by an ID, linked to a private/public key pair. The logical entity described is hosted on a \textit{validator client}, a separate entity with access to one or multiple validator private keys.

\Paragraph{Nodes} A validator client interacts with the rest of the network through a \textit{consensus node}, also referred to as node throughout. A node manages its own identity, consisting of a public/private key pair, linked to an IP address and port number. This information is shared through \textit{Ethereum Node Records (ENR)}, which are propagated across the network. Importantly, a node can host any number of validators, as these two entities are kept separate, in part for security reasons (see \Cref{sec:mitigation}). Moreover, a node might not host any validators, e.g., researchers who simply want to read the blockchain state.

\begin{figure}
    \centering
    \includegraphics[width=0.95\linewidth]{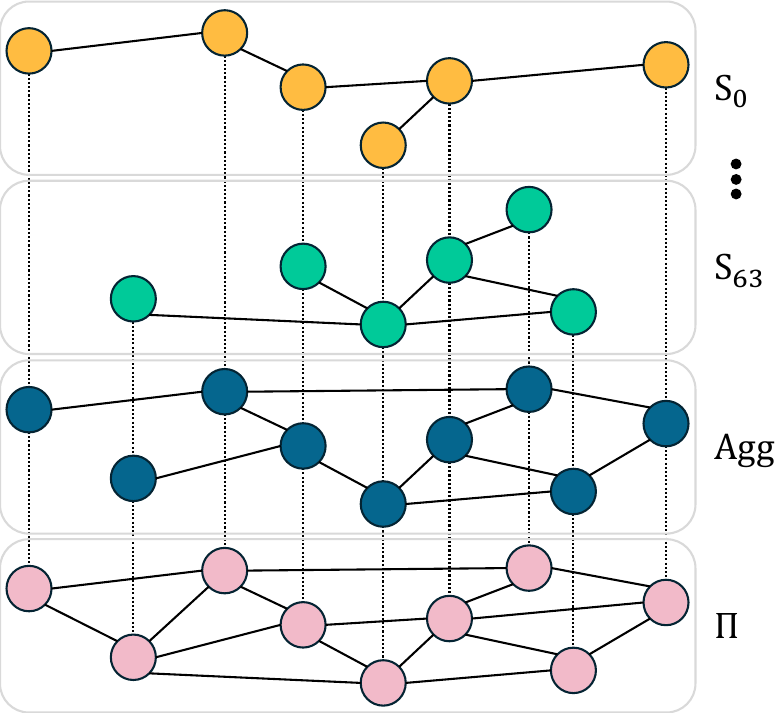}\vspace{-6pt}
    \caption{The graph $\Pi$ depicts the peering connections maintained by nodes. $Agg$ depicts the subgraph along which aggregations and blocks are propagated. It contains the same vertices as $\Pi$ with fewer edges. $S_0$ to $S_{63}$ depict the subnets and their associated subgraphs where attestations are propagated.
    By default, nodes randomly select two subnets to participate in.}
    \label{fig:subnet_explained}\vspace{-8pt}
\end{figure}

\subsection{Scaling Attestation Dispersal in Ethereum}
The Ethereum P2P network protocol facilitates the exchange of messages -- most importantly blocks, attestations, and attestation aggregations -- among peers, which is required for the blockchain to progress.

Given the sheer scale of Ethereum's validator set, with more than one million active validators~\cite{beacon}, it is not feasible for every validator to vote (i.e., broadcast an attestation) in every slot or to broadcast each attestation to all nodes in the network. This is particularly important since Ethereum strives to enable solo stakers to run their own nodes, necessitating low hardware and network requirements. Thus, voting is divided along multiple dimensions. 

\Paragraph{Time Division Across Slots} Each validator is tasked with attesting only once per epoch in a randomly assigned slot, which means, on average, once every 32 slots. As a result, only a fraction of validators vote in each slot. In turn, to achieve deterministic safety, block finalization occurs at the epoch level through a finality gadget~\cite{buterin2017casper}. This approach trades off latency to reduce message complexity.

\Paragraph{Network Division Across Committees} Validators' attestations are further divided into 64 \textit{committees}~\cite{dmarzb2013committess}. Within each committee, a set of validators (16 on average) is assigned as \textit{aggregators}. These aggregators collect and combine attestations into a single aggregated BLS signature\cite{ethereum2021bls}. Consequently, a node does not need to hear every individual attestation to stay synchronized but can rely on these aggregations. This division of attestations into committees is mirrored in the network layer, which is also divided into 64 attestation \textit{subnets} (also called \textit{topics}) and an additional subnet for attestation aggregates (see Figure~\ref{fig:subnet_explained})~\cite{dmarzb2013gossip}. Each committee is assigned to one of these 64 subnets, and the corresponding attestations are broadcast only within the respective subnet.

\Paragraph{Gossip Protocol}
A further network optimization to reduce message complexity is that messages are shared using a probabilistic broadcast implementation called GossipSub~\cite{vyzovitis2020gossipsub}. When a validator signs an attestation, the consensus node to which it is connected \textit{publishes} the attestation to the corresponding subnet by sending it to a subset of peers that are part of this subnet, these are called the node's \textit{fanout} for this subnet. Note that the sending consensus node itself does not need to be subscribed to this subnet, as committee assignments for validators change every epoch. 

To ensure stability within a subnet, each node statically subscribes to two topics by default, performing \textit{backbone} duties, which can be considered \textit{static subscriptions}. Additionally, if nodes need to receive messages from non-subscribed subnets (e.g., for an upcoming aggregation duty), they can request a \textit{dynamic subscription}. Within a subscribed subnet, nodes choose a subset of peers that are also in the subnet to share messages with. The choice of this subset is based on peer-performance.\footnote{Performance is based on a peer-scoring function which takes into account several factors including speed and quantity of information forwarded~\cite{vyzovitis2022gossipsub}. Specifics on the behavior of this score function is out of scope.} Nodes forward all messages they hear about within a subnet to these best-performing peers. These connections make up the subgraphs of Figure~\ref{fig:subnet_explained}.

\section{Threat Model}\label{sec:threatmodel}

In this work, we consider the P2P network of the Ethereum consensus layer, where consensus messages are broadcast. This system's main goal is reliable, efficient and low latency message delivery among a dynamic set of nodes hosting validators. As a secondary goal, source anonymity is also desired~\cite{ethresearch2020packetology}. Below we detail two main threats in the absence of source anonymity.

We consider an attacker aiming to create a wide-scale deanonymization, namely map validators to IP addresses of their hosting nodes. Importantly, since all validators have equal stake, no particular validator must be targeted.

We assume a weak, passive attacker, with access to a few nodes (without validators), limited time and hardware budget. The attacker only has access to exchanged network-level messages. All other data sources and attack vectors (off-chain data, clustering, etc.) are used for the verification of results.

Deanonymization is deemed successful if the probability that a given validator resides at a particular IP address is significantly higher than that expected from random assignment. The feasibility of deanonymizing a large portion of Ethereum's validators within the P2P network presents significant security concerns, enabling various types of attacks. These threats include those detailed in the following, as well as others, such as undermining assumptions critical for Danksharding~\cite{das2023a16z} (a new Ethereum upgrade) and compromising censorship resistance. Notably, even imperfect knowledge in these scenarios can inflict substantial harm on the network.

\subsection{Taking-Out Preceding Block Proposers} 

Ethereum block rewards comprise both consensus layer rewards and execution layer rewards. The consensus layer rewards are constant and independent of the transactions included in a block. In contrast, execution layer rewards consist of transaction fees paid for (preferential) block inclusion. On average, execution layer rewards are over three times higher than consensus layer rewards and can surpass them by several orders of magnitude due to the presence of \textit{MEV (Maximal Extractable Value) }in the Ethereum blockchain~\cite{HeimbachEthereum2023}. MEV is any value that can be extracted by the block builder by including, excluding, and reordering transactions in a block.

This disparity poses a risk to the consensus layer, as previous works~\cite{DaianFlash2020, QinQuantifying2022} have shown. Specifically, the consensus layer is vulnerable to \textit{time-bandit attacks}, where it can be rational for a block proposer to fork the blockchain and extract the MEV from earlier blocks for personal gain. The traditional time-bandit attack involves forking out the block created by the previous proposer, which is more challenging in Ethereum's PoS system than in PoW~\cite{2022reorg}.

Critically, attackers can exploit the lack of privacy in the P2P network to enhance time-bandit attacks and bypass the need to fork out the previous proposer's block.

An attacker, as the proposer of slot $n+1$, can prevent the proposer of slot $n$ from submitting a block proposal~\cite{hopr}. By doing so, the attacker could claim higher execution layer rewards, incorporating transactions from the previous slot and any new transactions arriving in the interim.

The attacker, as the proposer of slot $n+1$, knows the identity of their victim (the proposer of slot $n$) in the consensus layer well in advance, as proposers are assigned at least one epoch (approximately six minutes) beforehand. If the attacker can deanonymize the victim in the P2P network, they might temporarily sever the victim’s connection to the network through a (D)DoS attack or BGP hijacking. This disruption only needs to last four seconds, as block proposers have a four-second window to submit their proposal.

The success of this attack hinges on the attacker’s ability to both identify the victim in the P2P network and disconnect them. While some victims may have implemented mitigations (as outlined in Section~\ref{sec:mitigation}) that prevent the attack, an attacker can simply wait for another opportunity when they are selected as the block proposer and attempt the attack again.

\subsection{Breaking Liveness and Safety} 
\label{sec:threatmodel:breaking}
Even more concerning is the possibility of an attacker escalating the previous attack to compromise the liveness or safety of Ethereum. To disrupt liveness, an attacker could repeatedly sever the connection between the upcoming block proposer and the network. If every such attack succeeded, the blockchain would come to a standstill. However, achieving consistent success is improbable due to the mitigation measures discussed in \Cref{sec:mitigation}. Large staking pools, such as Lido, are particularly likely to implement robust defenses. Nevertheless, even if an attacker could disrupt just one in every ten proposers, the consequences for the blockchain would be significant, leading to delays and instability.

Rather than preventing a proposer from submitting a block entirely, the attacker might instead seek to disrupt block propagation to over one-third of the network. This could be achieved by launching a sustained DoS attack on a significant portion of the network or leveraging BGP hijacking. Such an attack would halt liveness, as the finality gadget -- responsible for confirming blocks -- would fail to gather the quorum required to finalize blocks. Again, the success of this strategy hinges on the attacker’s ability to prevent block proposals from reaching at least one-third of validators.

Alarmingly, an attacker could also threaten the safety of the blockchain. Many Ethereum light clients optimistically treat the chain's head as finalized. By breaking synchrony assumptions through targeted DoS attacks or BGP hijacking, the attacker could exploit this optimistic behavior to introduce safety violations~\cite{d2022goldfish}. For example, conflicting blocks might appear finalized to different segments of the network, resulting in inconsistencies that undermine the integrity of the chain.

Although such attacks are complex and resource-intensive, they are made more feasible by the deanonymization techniques discussed in the following. %

\section{Deanonymization Methodology}
\label{sec:methodology}
To ensure that their attack is as successful as possible, an attacker would aim to locate all validators running on a peer they are connected to, and thereby link validator IDs to their IP address. We show that focusing on attestation messages and their dissemination suffices for our deanonymization attack.

\subsection{Ideal Approach}\label{sec:ideal}

\begin{figure}[t!]
    \centering
    \includegraphics[width=\columnwidth]{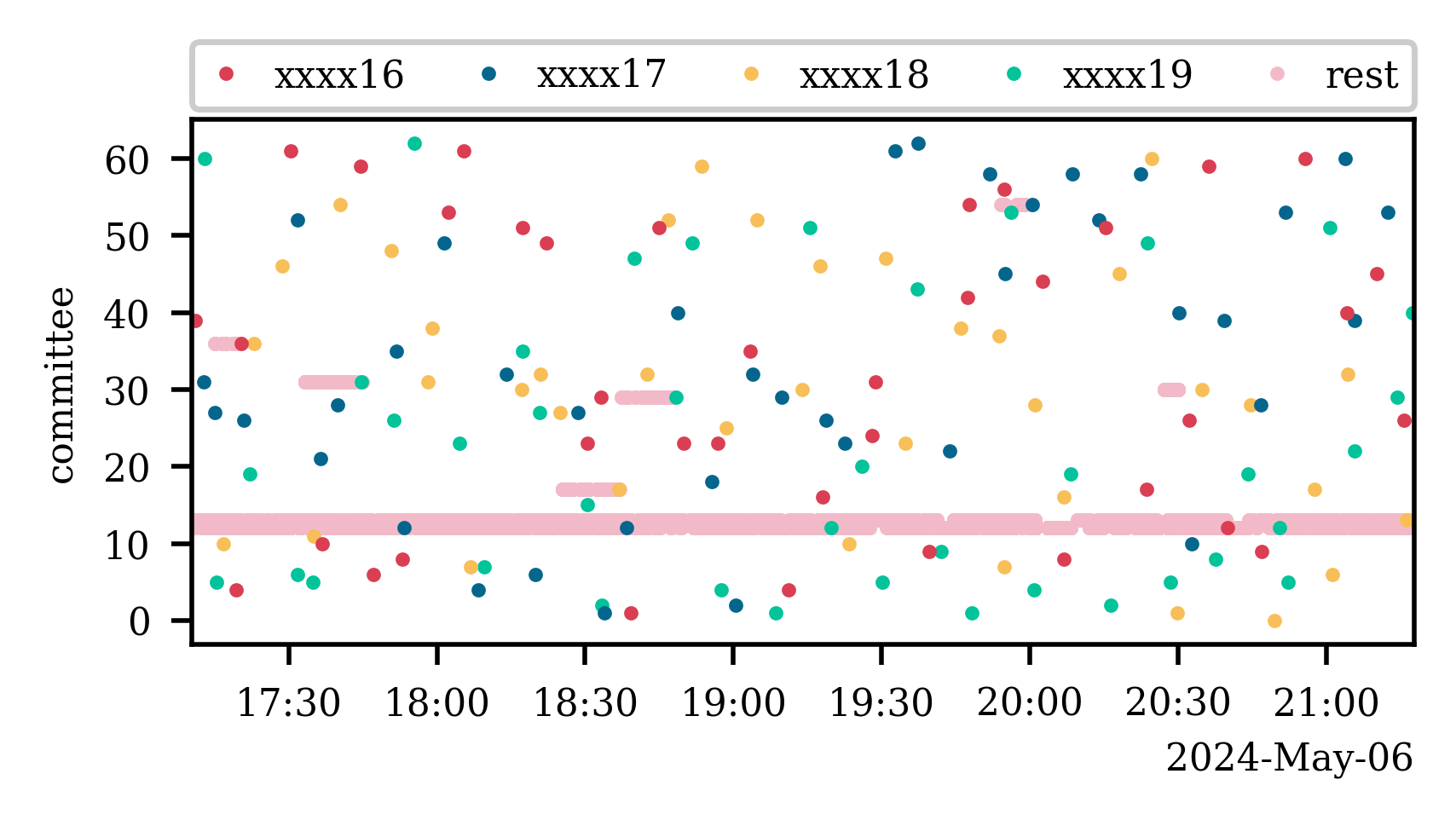}\vspace{-10pt}
    \caption{Attestations received from a peer over a four hour window, where the vertical axis corresponds to the 64 subnets. We identified four validators hosted on the machine, represented in red, blue, yellow, and green. The attestations by the remaining validators are shown in pink. Note that we have anonymized the full IDs of the identified validators, but emphasize that the IDs of these four validators are sequential. For the identified validators, we receive attestations from a wide variety of subnets. In contrast, the attestations from the remaining validators primarily come from the two subnets the peer predominantly serves (i.e., subnets 12 and 13), evidenced by the repeated flow of attestations for these subnets (see the long pink strip), as well as from short dynamic subscriptions (see the smaller pink horizontal strips).}
    \label{fig:casestudy}\vspace{-10pt}
\end{figure}

Given the background on Ethereum node behavior, we can describe how an ideal peer (a peer who gives us perfect information) would behave. We will guide the description with an example of a real peer we connected to in our experiments (see Figure~\ref{fig:casestudy}). Let us assume we are connected to a peer running $V$ validators who is a backbone in\footnote{We use the terms \textit{subscribed to} and \textit{backbone in} interchangeably.} two subnets. The peer's validators will thus attest $V$ times per epoch. Let us assume we receive perfect information from this peer, meaning we are in their fanout for all subnets and they forward all attestations they hear about in their two backbones to us. In each epoch, we will receive $V$ attestations from our peer for their validators, and $N \cdot \frac{2}{64}$ for all other $N$ validators.

\Paragraph{Observation} An ideal peer will only send us an attestation in a subnet they are not a backbone of if they are the signer of the attestation, and we are in their fanout for the corresponding subnet of the attestation.
\smallskip

Thus, in this scenario, we receive all attestations for the $V$ validators of the peer and can distinguish them as the only attestations we do not receive from the two backbones of the peer. Thus, linking validators to peers in this scenario is trivial. In practice, however, network message data is not perfect. 

\smallskip
\Paragraph{Imperfect Information in Practice}
To showcase this, we turn to the example peer in Figure~\ref{fig:casestudy}. On this peer, we will identify four validators associated with the peer;\footnote{Using our heuristics introduced in the following subsection.} their respective attestations are highlighted in red, blue, yellow, and green, while the remaining attestations are shown in pink. Notice that the attestations from these four validators, who have consecutive identifiers, appear equally distributed across subnets. In contrast, the vast majority of attestations come from the two subnets where the peer acts as a backbone (subnets 12 and 13 for the sample peer in Figure~\ref{fig:casestudy}). Thus, we can locate validators on our peers by observing how the attestations belonging to a validator, which we receive from the peer, are distributed across subnets. Additionally, and this is where the imperfect information comes into play, the validators hosted on the peer are occasionally tasked with being aggregators in a subnet approximately every 30 epochs per validator. During these times, they temporarily become a backbone (see the smaller pink horizontal strips) for these subnets and receive attestations from multiple validators belonging to the subnet.

In practice, five main factors impact the level of information we receive from a peer. First, nodes may run at non-default parameters, such as in more than two backbones, increasing the number of messages we receive from the peer. Second, we may not receive all attestations that a peer generates due to network reasons (e.g., disconnections or dropped network packets), or dynamic membership in a peer's fanouts which is out of our control (a peer chooses whether to include us in a fanout or not). Third, due to dynamic subscriptions or delayed information, we may receive some backbone attestations from a peer and not label them as such (see Figure~\ref{fig:casestudy}). Fourth, a validator client might use multiple nodes to propagate their messages or use different nodes for separate tasks~\cite{vouch2024}. Finally, fifth, we only get propagated attestations from peers if they give us a good peer score, thus we have to participate in attestation propagation. Us announcing a subset of attestations to our peers first in each backbone results in fewer attestations from each peer. This should be rare if they are the original signer of the attestations, and thus we make the following assumption to guide the development of our heuristics.

\Paragraph{Assumption} A peer will be the first to tell us about their attestation \textit{most} of the time. %

\subsection{Heuristic Approach}
\label{sec:method:heuristics}

To handle imperfect information, we develop a heuristics deanonymization approach to filter peers and narrow down on validator IDs as follows: we consider a validator $v$ to be hosted\footnote{As pointed out in Section~\ref{sec:ideal} a validator client might use multiple nodes to propagate their messages. Thus, to put it differently, the node is \textit{associated} with the validator and we will use the terms interchangeably throughout. We also observe such behavior, i.e., validators associated with multiple nodes, as we will outline in Section~\ref{sec:verification}. However, this behavior is rare and likely only used by node operators of large staking pools given that running a consensus node is expensive. Further, our methodology in theory allows us to locate all consensus nodes associated with a validator.} on a peer $p$, if \textbf{all} of the following conditions hold: 

\begin{itemize}[topsep=0pt,itemsep=-1ex,partopsep=1ex,parsep=1ex]
    \item [C1] The proportion of non-backbone attestations for validator $v$ exceed 
$$0.9 \cdot \left( \tfrac{64 -n_{\text{sub}}(p)}{64}\right),$$
where $n_{\text{sub}}(p)$ is the average number of subnets the peer is subscribed over the connection's duration. 
    \item [C2] The peer is not subscribed to all 64 subnets. 
    \item [C3] We receive at least every tenth attestation we expect for the validator $v$ from the peer.
    \item [C4] The number of attestations we receive for the validator $v$ from the peer $p$ exceeds the mean number of attestations per validator from peer $p$ by one standard deviation. 
\end{itemize}

By participating in all subnets w.h.p. we are added to several fanouts of our peer.
\footnote{We discuss this likelihood further in Appendix~\ref{app:heuristics}.}

Condition 1 ensures that the proportion of attestations we receive from $p$'s fanouts for a validator $v$ is at least 90\% the expected value. We choose 90\% conservatively to analyze only those peers from which we receive most attestations across all subnets.\footnote{In robustness-testing we find that if we lower this to 30\% we find an additional 10K validators overall, accounting for less than 4\% new validators.}
 
Condition 2 excludes any validators that are in all backbones as we would get no non-backbone attestations. 
Condition 3 ensures that we analyze the peers for which we are in several fanouts, removing those that would require a more careful analysis (we leave these peers unlabeled).\footnote{We use a conservative bound of at least one-tenth of messages across all subnets, though consistent attestations in just one fanout may already be enough to link validators.}
Lastly, condition 4 disregards peers that are participating in rare, non-default behavior such as broadcasting attestations in all subnets without advertising their subscriptions. 

The above conditions represent conservative heuristics such that we analyze only peers for which our connections were stable enough, for long enough, to ensure we have sufficient data. 
In Appendix~\ref{app:heuristics}, we provide a detailed explanation of all heuristics and robustness testing for each parameter choice.

\section{Data Collection}

We provide an in-detail explanation of how we collect data to both run our heuristic deanonymization approach and validate its results.

\subsection{Ethereum Network-Layer Logging}

We log all Ethereum network-layer messages by writing our own listening node implementation based on the Prysm client~\cite{prysmaticlabs2024prysm} (the most wide-spread consensus layer client for Ethereum~\cite{client2024prysm}). We call our modified consensus node \textsc{Rainbow} throughout, as \textsc{Rainbow} acts as a Prysm, taking peer input and breaking it down into the ``colors'' of the validators.

\textsc{Rainbow} connects to up to 1,000 peers, and statically subscribes to all subnets. These modifications allow us to deanonymize a larger set of nodes given that we have a higher peer count and observe attestations from all subnets. Other than that, it behaves as any Prysm node would, but logs three main sources of data. First and foremost, all attestations, their origin, and their origin subnet are logged. 
Second, this data is enriched by logging all advertised static subscriptions of our peers. Finally, we save precise connection data for all nodes we interact with.

Three instances of \textsc{Rainbow} are run alongside a Geth execution client on AWS \texttt{r5a.4xlarge} machines, in the \textit{us-east-1} (referred to as VA throughout), \textit{eu-central-1} (referred to as FR throughout), and \textit{ap-northeast-2} (referred to as SO throughout) data centers respectively. Additionally, we run one node on a bare-bones server in Zurich (referred to as ZH throughout). The data collection spans three days from (May 7, 2024 00:00 UTC to May 10, 2024 00:00 UTC). About 700 GB of compressed data is collected and loaded into SQL databases for further processing.

Due to the gossiping nature of the network layer, the same attestations are often received from multiple sources. In this work, we only consider the attestation that we receive first.

\subsection{Ethereum Network Coverage}\label{sec:crawler}
    
To understand our coverage of the Ethereum Beacon network, we gather measurements of all nodes in the network via our own \textit{crawler} implementation. 

Similarly to the execution network of Ethereum (detailed in \cite{kiffer2021under} and \cite{gao2019topology}), a node in the Beacon network keeps \textit{a peer table} 
of nodes they have heard about, which they reference when they need to make more peer connections. 
The peer table is based on the Kademlia hash table \cite{maymounkov2002kademlia} where each node stores ENRs\footnote{An ENR record encodes various information about a node including its network public key, IP, TCP/UDP ports, protocols supported and their versions, among others.} into buckets containing nodes whose ids (a 256-bit unique identifier computed from the ENR) has an XOR distance of $i$ from itself for $0\leq i <256$. Nodes periodically update their peer tables using the discv5 peer discovery protocol \cite{discv5}. We leverage the discovery protocol messages to enumerate a node's peer table, and repeat this for all nodes we hear about in the network.

\label{app:crawler}

 \begin{figure}[t]\vspace{0pt}
        \centering
        \includegraphics[width=\columnwidth]{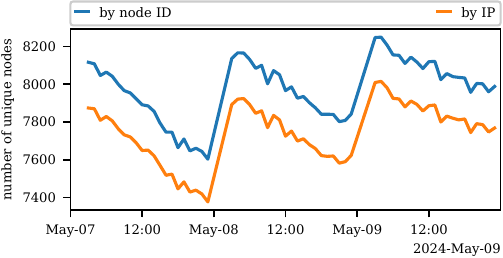}\vspace{-6pt}
        \caption{Reachable Beacon network nodes discovered in our crawls over our measurement period.}
        \label{fig:crawler_reachable}\vspace{-0pt}
\end{figure}
We run our crawler starting at 00:00 UTC on May 7, 8, and 9, and record all discovered  ENRs for nodes running on the main Beacon network. %
We discover an average of 16.5K unique IPs (20K unique node IDs) running on the mainnet each day, with a cumulative 20,240 IPs (28,998 IDs) found in the three days of crawls.

Additionally, each hour we send an in-protocol ping message to all ENRs discovered in the previous crawl to distinguish nodes that are reachable in the network (i.e., are not behind a NAT or firewall and thus can accept incoming connections from our \textsc{Rainbow} nodes), we show this in Figure \ref{fig:crawler_reachable}.\footnote{We note that the in-protocol ping messages use the node's ENR value, and some nodes do not respond if the ENR used is not up-to-date. Since the ENR changes if any of the values encoded change (e.g., a change in subnet subscriptions), ENRs change with some frequency. The gradual drop of responses over time we see in Figure \ref{fig:crawler_reachable} can be explained by both general network churn and ENRs learned in the previous crawl becoming outdated.}
We discover an average of 8.1K unique IPs (8.4K node IDs) that are reachable (at some point) each day, with a total of  8,941 IPs (9,468 IDs) of reachable nodes found in the three-day measurement period. 
    
Thus, we can lower-bound the number of reachable online nodes in the network during our measurement period as 8,941. Our \textsc{Rainbow} nodes are able to maintain sufficiently long connections with approximately half of these (see Section~\ref{sec:peeroverview}).

\subsection{Validator Entity Labeling}\label{sec:label}
To validate and verify our deanonymization, we perform a validator clustering, i.e., grouping Ethereum validators by the entity they belong to (e.g., staking pools). We then investigate whether the sets of validators located on the same machine are consistent with the validator clustering -- validators located on a peer belong to the same entity. 

We take the \texttt{pubkey\_mapping} dataset from \texttt{mevboost.pics}~\cite{mevboost_pics}, which provides labels for validators, and make the following amendments:

\begin{itemize}[topsep=0pt,itemsep=-1ex,partopsep=1ex,parsep=1ex]
    \item We monitor the Beacon chain deposit contract on the Ethereum blockchain to collect the \textit{deposit address(es)} -- the address(es) used to deposit the 32~ETH required for validator activation. For the top 15 entities, we identify deposit addresses used over 100 times and attribute any validator funded by these addresses to the corresponding entity.
    \item We apply a similar approach to the \textit{fee recipient address(es)}, which are used by validators to receive execution layer rewards. For the top 15 entities, we gather fee recipient addresses used over 100 times. Any validator exclusively using one such address is then attributed to the corresponding entity.
    \item Finally, we remove any labels for validators that were assigned multiple labels in the data set. 
\end{itemize}

\section{\textsc{Rainbow} Nodes Analysis} 
We begin by reviewing the number of peers we connect to, followed by an analysis of our deanonymization results for the validators on these peers.

\subsection{Peering Overview}\label{sec:peeroverview}
\begin{table}[t]
 \resizebox{1\columnwidth}{!}{
\begin{tabular}{lrrr}
\toprule
 & \multicolumn{3}{c}{peers} \\
 \midrule
 & seen & with established connections & with long connections \\
  \midrule
FR & 7,656 & 6,975 & 1,017 \\
SO & 7,816 & 7,122 & 1,142 \\
VA & 10,213 & 9,821 & 2,207 \\
ZH & 9,578 & 7,784 & 1,942 \\
\midrule
overall & 11,219 & 10,785 & 4,372 \\
\bottomrule
\end{tabular}
}\vspace{-6pt}
\caption{Number of unique peers seen, with connections and with long connections (i.e., $>$32 epochs) by each \textsc{Rainbow} node and overall. We consider a unique peer to be a unique IP port combination. VA and ZH \textsc{Rainbow} nodes saw and had more (long) connections than those in FR and SO.}\label{tab:connections}\vspace{-6pt}
\end{table}

Over the three-day data collection period, our four rainbow nodes attempted connections to 11,219 unique peers and successfully established connections with 10,785 (see Table~\ref{tab:connections}). Finally, they held sufficiently long connections with 4,372 unique peers over the four rainbow nodes, representing approximately half of the reachable network (see Section~\ref{sec:crawler}) and one-third of the estimated total network size~\cite{multi_node_explorer}.
Here, we consider a peer to have a long connection, if the total connection time over the three days lasted more than 32 epochs. Note that we disregard individual connections shorter than one epoch. Throughout, we consider a unique peer to be a unique IP port combination.\footnote{Note that we used unique IPs in the crawler as not all ENRs provide a port, but for peer connections, we may connect to more than one node behind an IP but at different ports. Additionally, a node may change its public peer ID, though the same validators remain at the node.}

We further note that the VA \textsc{Rainbow} node attempts connections and also successfully connects to the highest number of peers, whereas the ZH node only trails behind slightly. The FR and SO nodes exhibit smaller numbers of peers, both the number they attempt to connect and the number to which they establish long connections to. In particular, FR and SO only have around half the number of long connections in comparison to the other two \textsc{Rainbow} nodes.

\begin{figure}[t]
    \centering
    \includegraphics[scale = 1]{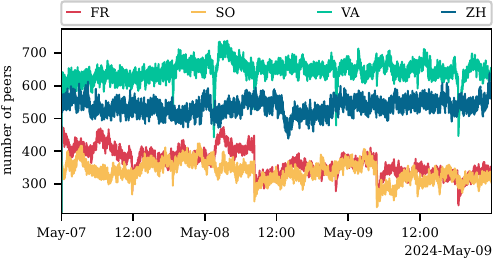}\vspace{-0pt}
    \caption{Number of peers each of our \textsc{Rainbow} nodes is connected to over time. On average, the VA node is connected to 645 peers, ZH is connected to 537 peers, FR is connected to 369 peers, and SO is connected to 338 peers.}\label{fig:number_peers}\vspace{-6pt}
\end{figure}

This is also evident when looking at the number of peers the nodes maintain at any point in time. In Figure~\ref{fig:number_peers}, we plot the number of peers our \textsc{Rainbow} nodes have throughout the measurement period. We observe that VA has the most peers on average with 645 and, except for a couple of short-lived dips in the peer count, maintains more peers than any other machine. Similarly, the ZH machine consistently has the second-highest peer count with an average of 537. Finally, the FR and SO \textsc{Rainbow} nodes have lower peer counts on average with 369 and 339 respectively. Further, their peer counts exhibit very similar patterns, i.e., dips in the peer count around the same time. For instance, around 8:00 UTC on 8 May. Interestingly, the peer count on the VA machine also drops around the same time, but immediately goes back up in contrast to the FR and SO \textsc{Rainbow} nodes where the peer count never fully recovers. We are unsure of the cause of these drops, this may be an artifact from AWS. However, while these drops in peer count potentially impact the number of peers we can deanonymize, they do not impact our accuracy. 

Lastly, we explore the unique peers we connected to for the period required for the analysis. We investigate the pairwise overlap across the \textsc{Rainbow} nodes in Table~\ref{tab:peer_overlap}. We observe the biggest relative overlap between the FR and SO \textsc{Rainbow} nodes with 496 -- around half of the peers these two nodes maintain long connections with overlap. Additionally, both the FR and SO nodes have an overlap of approximately 550 long peers with the VA node, while this constitutes more than half the long peers of the FR and SO nodes it only accounts for one-fourth of the VA long peers. Finally, the ZH node has the smallest overlap with the remaining nodes in relative and absolute terms. Possibly this is related to the three other nodes all being run from AWS and thus connecting to a more similar set of peers. %

\begin{table}[h]\vspace{-6pt}
\centering
 \resizebox{0.55\columnwidth}{!}{
\begin{tabular}{lrrrr}
\toprule
 & FR & SO & VA & ZH \\
\midrule
FR & 1,017 & 496 & 543 & 197 \\
SO & 496 & 1,142 & 592 & 205 \\
VA & 543 & 592 & 2,207 & 495 \\
ZH & 197 & 205 & 495 & 1,942 \\
\bottomrule
\end{tabular}
}\vspace{-6pt}
\caption{The pairwise overlap between the peers our \textsc{Rainbow} nodes (FR, SO, VA, and ZH) had sufficiently long ($>32$ epochs) connections with. In total, we had long enough connections to 4,372 peers over the three-day period.}\label{tab:peer_overlap}\vspace{-6pt}
\end{table}

\subsection{Deanonymization}
Next, we perform our deanonymization of the peers to which we were connected for a sufficiently long period. Using our \textit{Heuristic Approach} of Section~\ref{sec:method:heuristics} we divide peers into four categories as follows: 

\Paragraph{Deanonymized} We have located validator(s) on the machine with the aforementioned conditions. 

\Paragraph{No Validators} We did not receive a single non-backbone attestation from the peer and thus can safely assume that there are no validators hosted on the peer.\footnote{This gives us a conservative lower bound on the number of validator nodes, and roughly matches previous analyses which estimate about half the network to be running at least one validator\cite{brown2024exploring} - more details in \Cref{sec:relatedwork}.}

\Paragraph{64 Subnets} The peer is subscribed to all subnets. Thus, we will never receive a non-backbone attestation from the peer and our deanonymization does not work. Note that this is only a very small proportion of the peers. 

\Paragraph{Rest} Those peers from which we receive at least one non-backbone attestation but are not able to locate any validators hosted on the peer. Thus, these peers are in the grey -- there may or there may not be validators hosted on them. 

Multiple reasons could lead to a peer falling into the gray area. For one, we take a conservative approach to classifying validators, assuming we are in \textit{all} fanouts \textit{most} of the time for a given peer (i.e., receive 90\% of expected attestations). It is possible that we are only in a subset of fanouts or in some fanouts for short periods of time, thus receiving the peer's attestations for a relatively short period even though we maintain the connection for longer. Additionally, it could be that we received attestations that appear to be non-backbone from the peer before our \textsc{Rainbow} node was able to update the subnets a peer is subscribed to. 

In Table~\ref{tab:deanon}, we indicate the percentage of peers that fall in the respective categories. We start by noting that the distribution is similar across all three \textsc{Rainbow} nodes with the most significant chunk of peers either being deanonymized or hosting no validators. Further, only few peers subscribe to all 64 subnets. The remaining validators make up 7\% to 13\% depending on the \textsc{Rainbow} node, and less than 10\% overall.  

\begin{table}[t]
\centering
 \resizebox{0.9\columnwidth}{!}{
\begin{tabular}{lrrrr}
\toprule
 & deanonymized & no validators & 64 subnets & rest \\
\midrule
FR & 46.61\% & 43.91\% & 0.60\% & 8.88\% \\
SO & 43.75\% & 43.57\% & 0.26\% & 12.41\% \\
VA & 59.28\% & 33.12\% & 0.50\% & 7.10\% \\
ZH & 58.39\% & 33.35\% & 0.78\% & 7.48\% \\
\midrule
overall & 52.35\% & 37.52\% & 0.69\% & 9.46\% \\
\bottomrule
\end{tabular}
}\vspace{-6pt}
\caption{Percentage of peers that (1) we deanonymized validators on, (2) have no validators, (3) are subscribed to all 64 subnets, and (4) the rest by \textsc{Rainbow} node. Notice that the vast majority fall in the first two buckets, i.e., we can either locate validators on them or say with high certainty that there are no validators hosted on the peer.%
}\label{tab:deanon}\vspace{-6pt}
\end{table}

Note that the FR and SO nodes connect to a larger proportion of no-validator peers than the VA and ZH nodes. As a result, the VA and ZH nodes can locate more validators on their peers, i.e., deanonymize more peers. We hypothesize that VA and ZH have higher-quality information and thus maintain more long-term peers and make it to more fanouts. This is a chicken-or-the-egg scenario, where being better connected leads to better connections given the greedy behavior of the network. It is unclear why VA and ZH have this advantage. 

Finally, we wish to highlight that out of the peers for which we do not conclude that they host no validators, our methods can deanonymize 84.57\% across the four \textsc{Rainbow} nodes.

\subsection{Deanonymizations Over Time and Across \textsc{Rainbow} Nodes}

Next, we explore the impact of attack duration and the number of machines on the portion of the network that can be deanonymized.

\begin{figure}[t]
    \centering
    \includegraphics[scale=1]{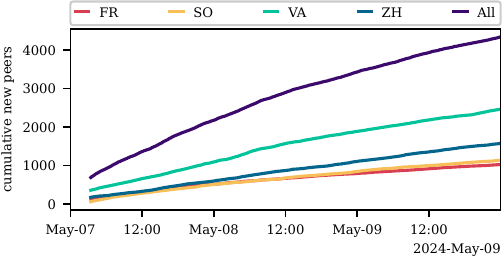}\vspace{-10pt}
    \caption{The cumulative number of peers each of our \textsc{Rainbow} nodes connected to over time, as well as the cumulative count across all.}
    \label{fig:cummulative_long_connections}\vspace{-6pt}
\end{figure}

In Figure~\ref{fig:cummulative_long_connections}, we look at the cumulative unique new peers our \textsc{Rainbow} nodes connected to with long connections over the measurement period.
We observe an almost linear increase in the cumulative number of new long connections, two-fold the amount for the VA machine with most new connections.  
This highlights the advantage of deploying multiple nodes to increase the number of validators located in the network.

\begin{figure}[t]
    \centering
    \includegraphics[scale=1]{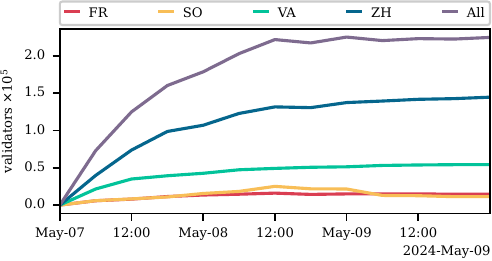}\vspace{-6pt}
    \caption{The number of validators (excl. service providers) deanonymized by our \textsc{Rainbow} nodes over time, as well as the cumulative count across all. Note that the number of deanonymized peers can decrease over time as an initially good connection to a peer can degrade over time. This is particularly evident for the SO \textsc{Rainbow} node.}
    \label{fig:validators_over_time}\vspace{-6pt}
\end{figure}

Additionally, we also look at the number of validators deanonymized by our nodes over time in Figure~\ref{fig:validators_over_time}. We perform our deanonymization on six hour fragments of our data, i.e., we analyze the first 6, 12, 18, and so forth, hours of data up until the entire collection period. Note that the number of deanonymized peers can decrease over time, which is particularly noticeable for the SO \textsc{Rainbow} node. This may occur if an initially strong connection to a peer weakens as time passes such that it no longer satisfies our heuristics.\footnote{Such behavior might be observed if a peer removes us from a fanout.}
While a more careful analysis could consider only the best connection period for each peer, for a more cautious analysis we omit these from the numbers in the previous section.

In Figure~\ref{fig:validators_over_time}, we observe a sharp initial increase in the number of deanonymized validators\footnote{Note that this figure excludes validators located on P2P service provider. See \Cref{sec:service} for details.} during the first one and a half days. Although this trend slows significantly over time, the total number of deanonymized validators continues to grow throughout the measurement period. This growth is smaller than that of new peer connections (see Figure~\ref{fig:cummulative_long_connections}) as we come across more nodes that are not running validators.

Overall, we conclude that various factors impact the success of deanonymization. A longer time horizon and deploying more nodes are generally advantageous, as highlighted in our analysis. Other factors, such as geographical location and hardware configurations (e.g., bare-metal servers versus virtual machines), significantly influence a node's performance within the network~\cite{kiffer2021under}. Low-latency connections in particular play a critical role in determining the performance of a node in the network. At the same time, networks are inherently complex, and a node's performance can vary unpredictably.

An attacker aiming to deanonymize the most validators would benefit from deploying multiple nodes in the network over an extended time horizon. The associated costs depend on how the nodes are deployed. Our measurements were relatively expensive, as we operated three nodes using AWS machines, with network traffic being the primary expense.\footnote{Each node incurred a cost of approximately 100~US\$ per day.} However, in general, a residential internet connection is sufficient, and the hardware requirements for Ethereum full nodes are modest~\cite{ethereum_node_requirements_2024} %
(less than 1,000~US\$). While these expenses are not trivial, they are minor compared to the potential profits from such attacks. For context, the total MEV extracted since the Ethereum Merge in September 2021 amounts to 598,019 ETH (approximately 2~billion~US\$)~\cite{mevboost_pics}. 

Furthermore, an attacker could optimize costs by refining their strategy -- for example, rotating through peers instead of keeping long-term connections to peers on which validators are already deanonymized.

\section{Verification}\label{sec:verification}
Overall, we locate 252,293 validators across the four \textsc{Rainbow} nodes -- approximately a fourth of the validator set (see column one in Table~\ref{tab:number_validators}). The ZH \textsc{Rainbow} node deanonymizes the majority with 215,293 while the VA node deanonymizes 74,904. Further, we deanonymize 13,771, and 14,388 validators with the FR and SO nodes respectively. 

\begin{table}[t]
\centering
 \resizebox{\columnwidth}{!}{
 
\begin{tabular}{lrrr}\toprule

 & validators & validators (excl. service providers) & non-unique validators\\
 \midrule

FR & 14,388 & 14,388 & 4,363 \\
SO & 13,771 & 11,185 & 2,411 \\
VA & 74,904 & 52,916 & 3,415 \\
ZH & 215,293 & 132,443 & 16,062 \\
\midrule
overall & 252,895 & 161,057 & 16,172 \\
\bottomrule

\end{tabular}
}\vspace{-6pt}
\caption{Number of validators located by each \textsc{Rainbow} node and overall. The first column indicates the total number of validators, the second column excludes validators from peers we identify to be P2P service providers (see Section~\ref{sec:service}), and the third column indicates the validators with a non-unique mapping to an IP port combination (see Section~\ref{sec:unique}). }\label{tab:number_validators}\vspace{-6pt}
\end{table}

Before delving deeper into the insights about the Ethereum P2P network derived from our deanonymization, we first validate the deanonymization itself. Notably, while we lack access to ground truth data, we employ several methods to assess the accuracy of our results. Although these methods are not definitive, they collectively provide valuable insights into the quality of our deanonymization. Specifically, we undertake the following three steps: (1) verify the consistency of validators hosted on the same machine, (2) analyze whether validators can be uniquely identified as being hosted on a single machine, and (3) confirm whether the same set of validators is identified on a peer across multiple \textsc{Rainbow} nodes.

\subsection{Consistency of Validator Sets}\label{sec:service}

We start by investigating whether the set of validators our analysis concludes to be hosted on each peer is consistent with their attributes, e.g., they all belong to the same staking pool as opposed to belonging to five different staking pools. We consider a set of validators on a peer\footnote{Recall that a peer is a unique IP port combination.} to be consistent, if one of the following holds: 
\begin{itemize}[topsep=0pt,itemsep=-1ex,partopsep=1ex,parsep=1ex]
    \item[$\Gamma1$] We have entity labels for at least 30\% of the validators and at least 90\% of these are identical.\footnote{Note that we make five exceptions where we find validators from two staking pools that both enlist at least one common node operator on the same machine. It appears that the node operator runs validators from different staking pools. We provide a discussion of this phenomenon in Section~\ref{sec:staking}.}
    \item[$\Gamma2$] At least 90\% of the validators were funded by the same deposit address.
    \item[$\Gamma3$] At least 90\% of the validators have exclusively used the same fee recipient address.
    \item[$\Gamma4$] The IDs of the validators can be aggregated into a few groups with consecutive IDs. We cap the number of groups to be at most one-tenth of the number of validators at that peer.
\end{itemize}
The first condition is the most straightforward -- we consider a validator set to be consistent if we have sufficiently many labels for the validator and they are largely identical. Note that we do not require them to be fully identical as there could be small inconsistencies in the entity label data set. Conditions 2 and 3 are very similar in that we require that the addresses used to deposit the stake and receive the execution layer rewards are largely identical for the validators in the set. Using the same address for either indicates that the validators belong to the same entity, i.e., the entity that controls the respective address. Finally, if the validators in the set have consecutive IDs this also indicates that they belong to the same entity, as the entity is likely to deposit the 32~ETH stake for a couple of validators in a row. This would otherwise be too much of a coincidence. Finally, a peer where we locate only a single validator is trivially consistent.

Next, we consider the validator set associated with a peer to be \textit{inconsistent} if: 
\begin{itemize}[topsep=0pt,itemsep=-1ex,partopsep=1ex,parsep=1ex]
    \item [C1] We have entity labels for at least 10\% of the validators, \textit{but} less than 90\% of these are identical (except if they are not all ENS names\footnote{Human-readable names associated with Ethereum addresses} or Rocketpool\footnote{Rocketpool validators are often home run and an individual could run their personal and Rocketpool validators on the same machine.}).
\end{itemize}
The above condition is the counterpart to condition 1 for labeling a validator set as consistent. Importantly, we are less restrictive, i.e., only require 10\% of the labels as opposed to 30\%, when deciding whether the validator set is inconsistent. We make exceptions for two known examples where we expect inconsistencies. All other peers are labeled as unknown.

Table~\ref{tab:label_consistency} shows the consistency of the validator sets located on the peers, both by percent of each node's peers and the proportion of located validators that are part of a consistent validator set. \textbf{We highlight that the vast majority of validator sets located on peers are consistent -- 93.75\% of peers across all machines.}

When we, however, turn to the proportion of validators these figures experience a significant shift -- only 63.67\% of validators located are part of consistent sets. This is a result of the inconsistent validator sets being significantly larger on average than the average validator set size. We proceed with a more in-depth investigation of these inconsistent validators.

\begin{table}[t]
\centering
 \resizebox{\columnwidth}{!}{
 
\begin{tabular}{l|rrr|rrr}
\toprule

 \multicolumn{1}{c}{}& \multicolumn{3}{c}{peers} & \multicolumn{3}{c}{validators} \\
 \midrule
 & consistent & inconsistent & unknown & consistent & inconsistent & unknown \\
 \midrule
FR & 95.50\% & 0.64\% & 3.85\% & 98.70\% & 0.28\% & 1.02\% \\
SO & 95.77\% & 0.60\% & 3.62\% & 79.89\% & 18.89\% & 1.22\% \\
VA & 94.08\% & 0.77\% & 5.15\% & 67.42\% & 31.02\% & 2.64\% \\
ZH & 91.74\% & 1.35\% & 6.91\% & 58.99\% & 46.63\% & 2.25\% \\
\midrule
overall & 93.75\% & 0.92\% & 5.33\% & 63.67\% & 39.67\% & 2.24\% \\
\bottomrule

\end{tabular}
}\vspace{-6pt}
\caption{Percentage of peers with a consistent validator set and validators belonging to a consistent validator set across our four \textsc{Rainbow} nodes as well as overall. Notice that the vast majority of peers have a consistent validator set.}\label{tab:label_consistency}\vspace{-6pt}
\end{table}

\begin{figure}[t]
    \centering
    \includegraphics[width=\columnwidth]{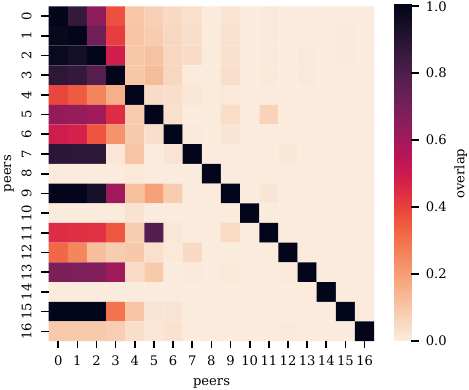}\vspace{-6pt}
    \caption{Pairwise overlap of the validator set of all peers with an inconsistent set of at least 200 validators. The darker the color of cell $(i,j)$, the larger the proportion of validators in the set of peer $j$ also found on peer $i$.}
    \label{fig:overlap}\vspace{-6pt}
\end{figure}

\Paragraph{Service providers} There are a total of 17 peers on which we locate inconsistent validator sets containing at least 200 validators, with the biggest validator set having 84,165 validators. We calculate the pairwise overlap between these sets and visualize the result in Figure~\ref{fig:overlap} with peers sorted by size (i.e., peer 0 has the largest validator set, and peer 16 has the smallest). Each peer has a non-zero overlap with at least one of the other peers, and there is significant overlap between many of the peers. We presume that these peers are what we will refer to as \textit{P2P service providers} throughout, i.e., service providers that help validators to quickly disseminate and receive messages e.g., bloXroute\cite{bloxroute}. 

Our main reasons for believing these peers to be service providers are that (1) they have access to attestations from a diverse (i.e., inconsistent) and large set of validators and (2) they have a large overlap in the validator set with different peers. Lower overlaps between these peers could be due to a variety of reasons including the peers belonging to different providers or geographical differences.

Given that we suspect these 17 peers to be P2P service providers, i.e., they likely do not operate the validators but have priority access to their messages, we exclude them from our deanonymization. The second column in Table~\ref{tab:number_validators} indicates the number of validators we locate after excluding the aforementioned service providers. \textbf{Overall we 161,057 validators -- more than 15\% of the network. }

\subsection{Uniqueness of Validator-IP Mapping}\label{sec:unique}

\begin{figure*}[ht!]\vspace{-0pt}

    \begin{subfigure}{\linewidth}
        \includegraphics[scale =1,right]{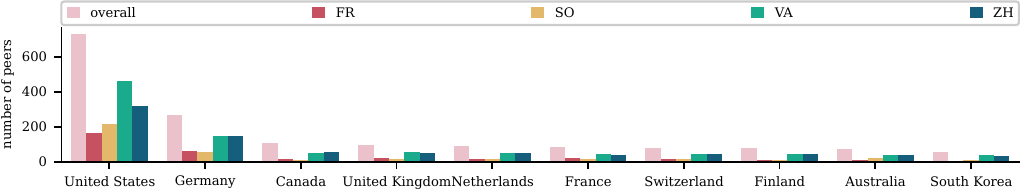}\vspace{-3.5pt}
        \caption{peers}\label{fig:country_peers}
    \end{subfigure}
    
    \begin{subfigure}{\linewidth}
        \includegraphics[scale =1,right]{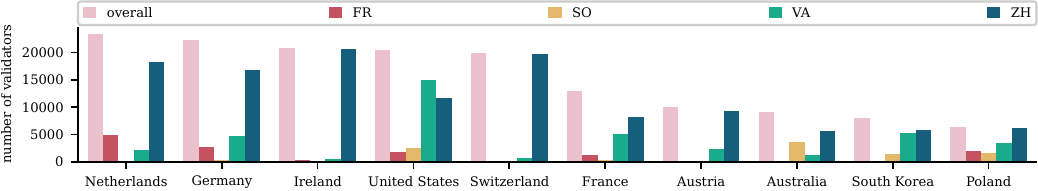}\vspace{-3.5pt}
        \caption{validators}\label{fig:country_validators}
    \end{subfigure}\vspace{-6pt}
    \caption{Number of peers deanonymized (see Figure~\ref{fig:country_peers}) and validators located (see Figure~\ref{fig:country_validators}) by top ten countries.}\label{fig:country}\vspace{-4pt}
\end{figure*}

To further verify our deanonymization we investigate the uniqueness of validator-IP port mappings -- whether we only map a validator to a single IP or multiple ones. Column three in Table~\ref{tab:number_validators} indicates the number of validators with non-unique IP mapping. In total, there are 16,172 such non-unique validators, roughly 10\% of all deanonymizations.

While this initially seems significant, closer inspection reveals special behavior by certain peers. Nearly three-fourths of the overlap involves peers in the same city, suggesting that some entities, possibly to ensure validator uptime, relay messages from their validators through multiple machines. As discussed in Section~\ref{sec:mitigation}, such practices can enhance validator privacy. However, our deanonymization still ties these validators to the peer sets we connect to. We find strong evidence of this behavior for many non-unique validator-IP mappings. The remaining overlap, less than 1\% of all deanonymized validators (excluding service providers), is minimal and may result from similar practices.

\subsection{Similarity of Deanonymizations}

The final analysis we perform is to verify our deanonymization is checking whether we locate the same set of validators on a peer (i.e., IP port combination) from multiple \textsc{Rainbow} nodes. In total, we deanonymized validators from more than one \textsc{Rainbow} node for 794 peers. For 762 (95.96\%) of these peers, we located the exact same set of validators on the machine from all \textsc{Rainbow} nodes that were connected to the peer. Further, on average the overlap of the validator sets is 99.20\%. \textbf{Thus, our deanonymization appears robust -- we locate the same validator set on a peer from different geographically distributed \textsc{Rainbow} nodes. }

\section{Insights}
\label{sec:insights}
We conclude our analysis by drawing insights from our deanonymization. Due to ethical reasons, we are careful to obfuscate any exact details -- we never reveal which validators we map to a particular IP or how a specific staking pool divides their validators across peers. We begin by investigating the pairwise overlap between validators we deanonymized across the four \textsc{Rainbow} nodes in Table~\ref{tab:number_validators_overlap} and find that there is quite a significant overlap between the validators deanonymized. For instance, the overlap between the validators deanonymized on the FR, SO, and VA \textsc{Rainbow} nodes with the ZH \textsc{Rainbow} node is more than half of the total validators deanonymized on the three former \textsc{Rainbow} nodes. 

\begin{table}[h]
\centering
\resizebox{0.7\columnwidth}{!}{ 
\begin{tabular}{lrrrrr}
\toprule
 & FR & SO & VA & ZH & unique \\
\midrule
FR & 14,388 & 2,842 & 5,326 & 10,769 & 2,675 \\
SO & 2,842 & 11,185 & 5,855 & 9,221 & 691 \\
VA & 5,326 & 5,855 & 52,916 & 27,717 & 23,577 \\
ZH & 10,769 & 9,221 & 27,717 & 132,443 & 93,854 \\
\bottomrule
\end{tabular}

}\vspace{-6pt}
\caption{Pairwise overlap of deanonymized validators across machines, as well as the number of unique validators per machine. With the exception of the ZH node, half of the validators deanonymized on a \textsc{Rainbow} node have also been deanonymized by at least one other \textsc{Rainbow} node.}\label{tab:number_validators_overlap}\vspace{-8pt}

\end{table}

\subsection{Geographical Distribution}

\begin{figure*}[t]\vspace{-2pt}
    \begin{subfigure}{\linewidth}
        \includegraphics[scale =1,right]{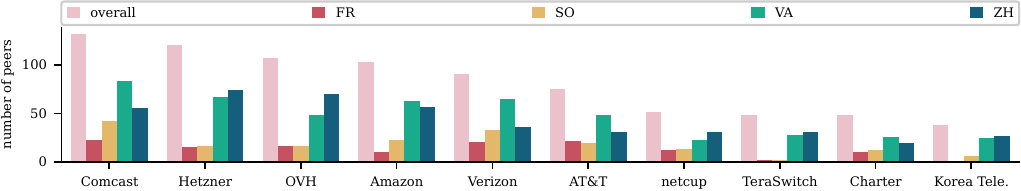}\vspace{-3.5pt}
        \caption{peers}\label{fig:org_peers}
    \end{subfigure} 
    
    \begin{subfigure}{\linewidth}
        \includegraphics[scale =1,right]{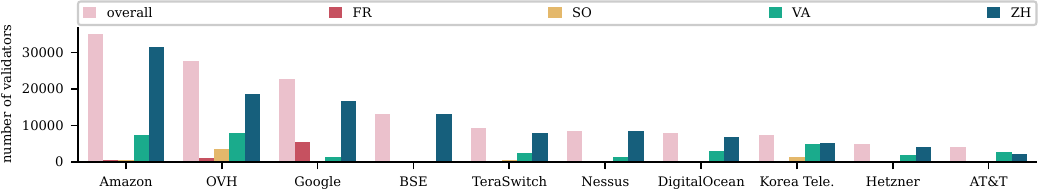}\vspace{-3.5pt}
        \caption{validators}\label{fig:org_validators}
    \end{subfigure}\vspace{-6pt}
    \caption{Number of peers deanonymized (see Figure~\ref{fig:org_peers}) and validators located (see Figure~\ref{fig:org_validators}) by top ten organizations. Note that while half of the deanonymized peers are run on cloud services, we locate $\sim$90\% of validators on cloud services.}\label{fig:org}\vspace{-6pt}
\end{figure*}

We continue by analyzing the geographical distribution of the peers we deanonymize. This not only gives insight into the network but can also help us understand the differences between the efficacy of our \textsc{Rainbow} nodes.

On a continent basis, 46.79\% of the peers we deanonymize are located in Europe, 38.05\% in North America, 9.95\% in Asia, 4.16\% in Oceania, 0.77\% in South America, and 0.27\% in Africa. In Figure~\ref{fig:country_peers}, we show the number of peers on which we deanonymized validators by \textsc{Rainbow} node as well as overall per country for the top ten countries. We deanonymized the largest number of peers in the United States making up 33.03\% of the peers. This is more than double the number of peers we deanonymize in the next two biggest countries. This is similar to the overall geographical distribution of peers in the network, where the United States and Germany make up the biggest proportions. Additionally, we also observe a geographical bias of our \textsc{Rainbow} nodes in terms of the peers on which they deanonymize. Most notably, the VA \textsc{Rainbow} node deanonymizes more peers in the United States than the ZH \textsc{Rainbow} node. Similarly, the ZH node deanonymizes more peers in most European countries in comparison to the VA node. 

In Figure~\ref{fig:country_validators} we further provide insights regarding the number of validators deanonymized by each \textsc{Rainbow} node per country. We start by highlighting the most striking difference to the previous figure: most validators we locate are hosted in the Netherlands (i.e., 12.71\%), while the United States only comes in fourth place. Note that a difference in the figures is expected given a non-uniform distribution of validators across machines. In general, we observe that there is a larger proportion of validators in Europe as opposed to peers with validators. In particular, 70.90\% of the validators we locate are in Europe, 12.48\% are in North America, 11.46\% are in Asia, 5.08\% are in Oceania, 0.05\% are in Africa and 0.03\% are in South America. Additionally, we again notice the same geographical bias, i.e., the SO node's high relative proportion of deanonymizations in Australia and South Korea.

We further repeat the preceding analysis for organizations instead of countries in Figure~\ref{fig:org}. The largest organization in terms of the number of peers is Comcast a residential ISP with 5.97\% and the following three (Hetzner, OVH, and Amazon) are cloud providers (see Figure~\ref{fig:org_peers}). Around half of the peers we deanonymize are hosted through cloud providers, whereas the other half run through residential ISPs, i.e., home stakers. 

There is a large shift in these figures when we turn from peers (see Figure~\ref{fig:org_peers}) to validators (see Figure~\ref{fig:org_validators}). For one, eight out of the ten largest organizations are cloud providers and all of the biggest seven are cloud providers. We locate the largest number of validators in Amazon data centers, i.e., 19.07\%. Surprisingly, the ZH node deanonymizes by far the largest proportion of validators in Amazon data centers \textbf{even though it is the only \textsc{Rainbow} node not run from an Amazon data center}. Finally, we highlight that around 90\% of the validators are run through cloud providers, with the other 10\% belonging to residential ISPs.

\begin{figure}[t]
    \centering
    \includegraphics[width=\columnwidth]{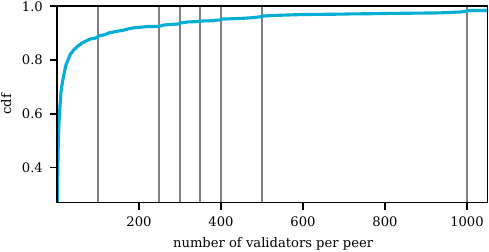}\vspace{-6pt}
    \caption{Cumulative distribution function (cdf) of the number of validators hosted per peer. We only locate a single validator on 27\% of peers, while there at least 100 validators located on around 11\% of peers. The vertical lines indicate 100, 250, 300, 350, 400, 500, and 1,000.}
    \label{fig:cdf}\vspace{-6pt}
\end{figure}

\subsection{Validator Distribution}

For those peers on which we deanonymize validators, we proceed by analyzing how many validators they host (see Figure~\ref{fig:cdf}).
There is only a small number of validators located on most peers: 
a single validator on around 27\% of our peers, with more than half of our peers having no more than four validators. In fact, we see that only 11\% of peers have more than 100 validators. We further note that an in-detail inspection of Figure~\ref{fig:cdf} reveals that for those peers that host many validators (i.e., more than 100), there is a bias towards a ``round'' number of validators to be hosted on a peer. This is evident by increases in the cdf at 100, 250, 300, 350, 400, 500, and 1,000 validators per peer. The gray vertical lines indicate these. For one, this indicates that large organizations running multiple validators on a machine tend to divide validators across nodes in round numbers as one would expect, e.g. an organization controlling 10,000 validators might divide them across 10 nodes and run 1,000 validators on each node. \textbf{Additionally, this observation also indicates that our deanonymization technique locates all validators hosted on a machine as it is less likely for these ``round'' numbers to be a coincidence. }

\subsection{Staking Pools}\label{sec:staking}

We continue our analysis by taking a deeper look at the practices of large staking pools, which we can observe as a result of our deanonymization. All statistics contained in this section concern the largest five staking pools (i.e., Lido, Coinbase, Ether.Fi, Binance, and Kraken~\cite{beacon}) and we do not reveal to which particular staking pool any statistic concerns. 

\begin{figure}[t]
    \centering
    \includegraphics[scale =.95]{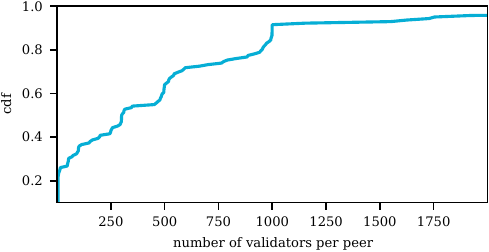}\vspace{-6pt}
    \caption{Cumulative distribution function of the number of validators hosted per peer for the biggest five staking pools.}\vspace{-6pt}
    \label{fig:big_cdf}
\end{figure}

In Figure~\ref{fig:big_cdf}, we plot a cdf of the number of validators hosted per peer for these five staking pools. Recall in our earlier analysis (see Figure~\ref{fig:cdf}) the observation that node operators of staking pools tend to run a ``round'' number of validators on the same machine. We observe these same increases in Figure~\ref{fig:big_cdf}. Here we observe an average of 709 validators on a given peer, with the larger validator set containing 19,390 validators. We highlight the security concern this raises: though there are a million validators, just hundreds of these peers going offline can stall Ethereum as more than two-thirds of the validators must be online to guarantee liveness. 

\begin{figure}[t]
    \centering
    \includegraphics[scale=1]{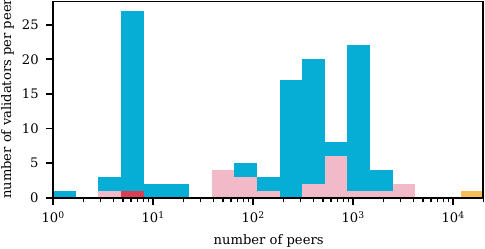}\vspace{-4pt}
    \caption{Number of validators per peer for the biggest five staking pools. Notice that we only deanonymize validators from four of the five biggest staking pools.}

    \label{fig:big}\vspace{-4pt}
\end{figure}

In Figure~\ref{fig:big} we examine the distribution of validators in large liquid staking services in greater detail. The histogram shows the number of validators hosted by peers from the top five staking pools (four of which are deanonymized). While some peers host fewer than 100 validators, most peers in these pools host over 100, highlighting the economic benefits of staking pools. The most validators we locate on a single peer is 19,390.

Additionally, many staking pools utilize node operators, and as a result, the protocols claim that the validators operated by the different node operators are independent, which would improve the decentralization of the consensus layer. However, given that many of the node operators run validators for various staking pools, this creates a dependency between the protocols. Even more so, we find five instances of validators from two different staking pools that utilize the same node operators being located on the same machine. Thus, many of the major staking pools are not (fully) independent of each other. This is especially concerning as the biggest staking pool alone already holds nearly a third of the staking power. Further, data regarding the distribution of validators across node providers is not easily accessible, thus making the decentralization of the Ethereum blockchain difficult to assess.

\subsection{Implications on Decentralization} 
Decentralization of the P2P network as well as Ethereum's consensus layer is essential to prevent one party from maliciously overtaking the network. The decentralization of Ethereum's PoS consensus layer is a topic of frequent discussion, given that the largest staking pool Lido controls nearly a third of the staking power -- a critical safety threshold in Ethereum PoS. To combat the criticism, Lido proclaims that their validator set is operated by multiple independent node operators and thus they should all be seen as independent entities. While it is unclear whether this characterization is accurate~\cite{grandjean2024ethereum}, our insights into the P2P network reveal a further entanglement between different staking pools. As we outline in the previous section, not only do node operators run validators for different staking pools but we even find that they run them from the same nodes. \textbf{This shows that staking pools are not always independent from each other, e.g., whenever such a node goes offline, validators from both protocols go offline.} Similarly, we find that some peers host thousands of validators. Our insights gained from the deanonymization raise questions regarding the decentralization and resilience of the Ethereum consensus layer.

\section{Mitigations}\label{sec:mitigation}
We now discuss mitigations against our proposed deanonymization technique presented in \Cref{sec:methodology}. 

\subsection{Providing Anonymity}
We begin by discussing techniques that enhance anonymity within the P2P network. Before delving into the details, it is important to note that the current max effective balance for a validator is 32~ETH, which results in many redundant validators being run and controlled by the same entity. However, in the upcoming Pectra hardfork~\cite{bankless2024pectra}, the max effective balance is set to increase to 2048~ETH~\cite{eip7251}. This change will allow entities to consolidate up to 64 validators into a single entity. Consequently, the next hard fork has the potential to significantly reduce the number of attestations propagated through the P2P network. This reduction allows for the following approaches that slightly increase the number of P2P messages to preserve anonymity. Importantly, with such an increase in P2P messages, the reductions in signature verification (i.e., one per validator per attestation) remain preserved.

\Paragraph{Additional Subnets} Taking over backbone duty for more than two subnets is a rather straightforward way to mitigate the uncovered flaws. Indeed, our methodology fails at deanonymizing nodes subscribed to 64 subnets. Raising the number of subnets, however, is in direct contrast to the very reason behind the existence of subnets: the reduction of the message complexity, from $ \sim n_{\text{nodes}} \cdot n_{\text{validators}} \cdot n_{\text{meshPeers}}$, by a factor of $\sim \frac{\text{avgSubscribedSubnets}}{\text{totalSubnets}}$. However, as outlined previously, the number of messages in the P2P network is expected to decrease with the upcoming hard fork, and thus approaches that (slightly) increase the average number of subscribed subnets should be tolerable. 

\Paragraph{Additional Nodes} Validator clients can connect to multiple nodes to defend against our deanonymization. This can be done through validator clients that support multiple connections (see e.g., the Vouch validator client~\cite{vouch2024}).

Validators have three main tasks to perform: broadcasting attestations, aggregating attestations, and proposing blocks (see \Cref{sec:background}). Our methodology solely relies on attestations, but, as discussed in \Cref{sec:threatmodel} the most straightforward implications of the deanonymization stem from hindering block proposals. Thus, a simple mitigation is to run two nodes, one for propagating and aggregating attestations, and another where block proposals are broadcast. Alternatively, an entity might decide to propagate their attestations through multiple nodes, to either (i) increase the hardware requirements for a DoS attack, or (ii) increase the time to deanonymization by capping the amount of attestation sent by any node.  

Solutions based on adding more nodes do not tackle the root cause of the issue (likely just increasing the complexity of deanonymization) and almost universally lead to higher message complexity. Furthermore, they also result in larger operating costs, that are especially difficult to bear for solo stakers, thus going against the ethos of Ethereum.

\Paragraph{Private Peering Agreements} Out of the box, the Lighthouse and Prysm Ethereum clients provide the capacity to perform private peering agreements -- a set of trusted peers can be defined who function as additional relays for gossip messages. This functionality is meant for increased performance and reliability, however, it also makes it harder to uniquely map validators to one IP. To be precise, through our presented deanonymization methodology, we could still determine the set of validators and the set of nodes that are part of the peering agreement, but not which validator is hosted by which node. Thus, a private peering agreement can provide \textit{k-anonymity}\footnote{In this context, a node is said to be k-anonymous if it cannot be distinguished from k-1 other nodes.} As a result, it would be necessary to target the $k$ peers in the peering agreement during an attack. This mitigation comes with two main caveats: First, $k$ has to be chosen large enough to make DoS attacks prohibitively costly. Second, while peering agreements can be established off-band, finding trustworthy peers may again be a prohibitive cost for small participants, with the introduction of such complexities potentially contributing to centralization. Based on our results presented in \Cref{sec:insights}, 50\% of nodes hosting validators host less than four.
Nonetheless, private peering agreements can at least mitigate against the threat of network-wide disruptions (as outlined in \Cref{sec:threatmodel}). The positive effect of such a mitigation can be seen today in validators who make use of service providers at the gossip layer (see \Cref{sec:verification}).

\Paragraph{Anonymous Gossiping} Established protocols like Dandelion~\cite{bojja2017dandelion,fanti2018dandelion++} and Tor~\cite{dingledine2004tor} have been considered in the past in the context of Ethereum~\cite{ethresearch2022dandelion,das2022tweet,andras2023tor}. The core idea of Dandelion is to first propagate messages along a path of a single node (the \textit{stem} phase), before releasing a message (the \textit{fluff} phase). This comes at a significant latency cost and is deemed incompatible with the economic incentives of publishing messages quickly~\cite{ethresearch2022dandelion}.

\subsection{Defending Against DoS} Both main threats outlined in \Cref{sec:threatmodel} can be exercised by employing DoS. A (partial) mitigation could thus consist of defending against DoS attacks. We consider this defense independently, as it is typically orthogonal to anonymity.

\Paragraph{Network Layer Defenses} The creators of libp2p provide a list of implemented mitigations~\cite{libp2p2020dosmitigation}, including limiting maintained connections, rate-limiting incoming connections, and automatically adjusting the firewall. Nonetheless, manual intervention may still be necessary, as these measures are not expected to fully prevent an attack.

Giuliari et al.~\cite{giuliari2024dos} present a list of traditional defense strategies, such as IP-based filtering, cloud-based DoS protection, overprovisioning, and VPNs. The authors argue that none of these are satisfactory for consensus algorithms in a blockchain setting. For Ethereum specifically, these concerns are even more insurmountable, due to (i) Ethereum's prohibitively large network size (see \Cref{app:crawler}) and (ii) the comparatively lower assumed economic power and hardware specifications of solo stakers. Giuliari et al.~\cite{giuliari2024dos} instead propose a defense based on \textit{source authentication and rate limiting}. While promising, source authentication represents a shift for Ethereum, and would thus require significant changes.

\Paragraph{Secret Leader Election} Another way to combat DoS attacks is for leaders (block producers) to remain anonymous until they perform their duties. This is the goal of secret leader election protocols~\cite{boneh2020single}. This approach does not conflict with any other assumptions needed, and might thus be the most promising approach. Multiple proposals have already been discussed in the context of Ethereum~\cite{ethresearch2022whisk, buterin2022secretelection}, but none have yet progressed beyond the design stage.

\Paragraph{Distributed Validator Technology (DVT)} By splitting the private key of a validator into multiple shares, using DVT a validator can run from multiple clients and create signatures even if some clients are unavailable~\cite{ethereum2024dvt}. The resilience gained may also help validators create blocks even when experiencing a DoS attack.

\section{Related Work} \label{sec:relatedwork}
 
    \Paragraph{Network Measurements and Attacks} 
    The measurements of Ethereum and other cryptocurrency networks most relevant for our work are those that measure gossip properties of these networks, observed miner centralization in Bitcoin and Ethereum~\cite{decker2013information,gencer2018decentralization,lin2021measuring,kiffer2021under}. Measurements of validator centralization at a network-level have been limited. In a previous protocol version of Ethereum~\cite{eipsubnets}, nodes in the network would subscribe to one (uniform) random attestation backbone per validator. Thus, previously, one could estimate lower bounds for how many validators were on a node. This was done in~\cite{brown2024exploring} where the authors estimated an average of 12K nodes in the network with 5-6K running at least one validator. 

    A large body of literature explores deanonymization attacks in the context of anonymity-preserving distributed systems, such as Crowds and Tor~\cite{basyoni2020survey}. Foundational models~\cite{wright2004predecessor,figueiredo2004OnTA} demonstrate how repeated path observations degrade anonymity over time, exploiting statistical biases towards identifying initiators (so-called predecessor attacks). Crowds' probabilistic routing amplifies this vulnerability compared to the fixed circuits used in Tor~\cite{panchenko2007crowds}.
    Other seminal works focus on traffic correlation~\cite{murdoch2005low}, timing attacks~\cite{shmatikov2006timing}, and learning-based attacks~\cite{nasr2018deepcorr}. In comparison, our work exploits the performance improvements introduced in GossipSub, making our deanonymization attack more effective.

    For cryptocurrencies specifically, deanonymization attacks have primarily taken advantage of network timing information, e.g., by showing the possibility of tracking the source of transaction data to the peer first to distribute the transaction in the Bitcoin~\cite{fanti2017deanonymization,biryukov2014deanonymisation,biryukov2019deanonymization} and its Lightning Network~\cite{rohrer2020counting}, as well as the Ethereum network~\cite{tang2023strategic,hopr}. Gossip protocols to prevent such deanonymization attacks have also been proposed, such as the family of Dandelion protocols~\cite{bojja2017dandelion,fanti2018dandelion++}. We highlight that the aforementioned timing-based deanonymization attacks require a connection to a large portion of the network to be sure of the originating source of messages for each address. Our attack, in contrast, takes advantage of protocol-level details that can be run from a single node to any number of its peers.
    The danger of the potential deanonymization of validators has been a topic of discussion on Ethereum for many years. However, to the best of our knowledge, apart from a meta-data-related technique~\cite{ethresearch2020packetology,github2018brainstorming}, no concrete deanonymization methodologies have been proposed. 

    There is also a broader history of network-layer attacks in cryptocurrency networks including Eclipsing~\cite{heilman2015eclipse,marcus2018low,henningsen2019eclipsing}, network-partitioning~\cite{tran2020stealthier,tran2021routing}, and routing manipulations such as BGP-Hijacking~\cite{apostolaki2017hijacking} and its prevention \cite{apostolaki2018sabre} and miner-pool networks routing attacks \cite{tran2024routing}. Protocol-level details have also been utilized to infer peering relationships, such as the structure of transaction messages and their broadcast behavior in the Bitcoin~\cite{neudecker2016timing,delgado2019txprobe,miller2015discovering} and Ethereum~\cite{li2021toposhot} networks, and peer discovery messages in the Monero network~\cite{cao2020exploring}. 

    \Paragraph{Gossip Networks} GossipSub was introduced in 2020~\cite{vyzovitis2020gossipsub}. An improved version (v.1.1) has been introduced a year later~\cite{vyzovitis2022gossipsub}.
    We are not the first to present a weakness of GossipSub. Kumar et al.~\cite{kumar2023formal} for instance present a formal method analysis of GossipSub and show that the score function is fair, albeit exploitable. In a companion paper~\cite{kumar2023verification}, they go on to synthesize and simulate attacks on GossipSub.
    
    Guerraoui et al.~\cite{guerraoui2023inherent} study the fundamental limit of achievable source anonymity in gossip on general graphs. Our analysis is different, as it mainly considers the effect on anonymity of splitting a single gossip network into multiple components.

\section{Outlook}
In this work, we present a simple and low-cost attack that allows a node in the network to deanonymize the validators associated with its peers. And to the best of our knowledge, we are the first to quantify the resulting lack of privacy in Ethereum's P2P network. 

There are several ways to strengthen the attack and deanonymize a larger portion of the network. Our \textsc{Rainbow} nodes ran with closed ports, limiting connections to the reachable part of the network. Running them with open ports would expand connections and likely deanonymize more validators. Additionally, we only consider the first occurence of each attestation, but including duplicates would further improve deanonymization. Lastly, our method focuses solely on attestations, but other validator tasks (e.g., participation in aggregation and sync committees) also leak information. This information could not only enhance deanonymization, but potentially also identify aggregators before they self-reveal, opening up additional attack vectors.

Given that our ``simple'' attack deanonymizes a large proportion of the validator set, the possibility of making it stronger in light of the outlined implications is evermore concerning. Thus, it is our belief that the development of a privacy-preserving mechanism for the Ethereum P2P network is of utmost importance, and we hope that our discussion of mitigations can guide future work.  

While this work focuses on Ethereum, it is important to emphasize that the foundational elements enabling the deanonymization attack are inherent to its use of the GossipSub~\cite{gossipsubspecs} protocol. Developers planning to integrate GossipSub into their systems should carefully consider the privacy implications associated with its subnet overlay feature.

\section{Ethics Considerations}
\label{sec:ethics}

Our research aims to explore the lack of privacy for Ethereum validators. As this research involves critical infrastructure and the identification of organizational entities that may not wish to be identified, we followed rigorous ethical guidelines in accordance with the principles outlined in the Menlo Report~\cite{bailey2012menlo}. Below, we assess key stakeholders and detail how each principle was addressed throughout the course of this study. 

\Paragraph{Respect for Persons} The principle of respect for persons requires that individuals' autonomy and privacy are protected, and informed consent is obtained whenever necessary. Validators are logical entities identified through pseudonyms, which by design should make it so that they cannot be linked to humans. Nonetheless, we ensured that no identifiable information about any entities running Ethereum validators was exposed. We minimized the use of identifiable data, by using it purely to verify the validity of our results. Moreover, the analysis was done using publicly available data from the Ethereum P2P protocols.

\Paragraph{Beneficience} We systematically assess risks by identifying key stakeholders, in order to maximize probable benefits while minimizing probable harm. 
First, staking entities that run Ethereum validators might suffer monetary harm, if an attacker were to exploit the vulnerability we expose. In order to minimize the likelihood of this scenario, we publish neither our code nor the deanonymized validator dataset (see Section~\ref{sec:openscience}). We further reached out to some larger entities, informing them of our findings. 
We were also careful to only run standard Ethereum clients to collect data. In particular, we avoided rewriting client code to make deanonymization more efficient, as this might have caused additional load on current Ethereum nodes. Instead, our clients most likely had a net positive impact, by contributing in forwarding P2P messages for the duration of our experiment.

The risk that malicious actors benefit from our disclosure is also counter-balanced by the mitigation strategies we put forward. Due to the improved understanding of privacy leaks, we can provide tools and mitigation strategies for Ethereum validators to minimize the efficacy of potential attacks.

The Ethereum Foundation could itself be harmed, as their reputation as stewards of the ecosystem might be questioned. To mitigate such negative consequences, we have been in contact with the Ethereum Foundation from the very start of the project. Once our study was finished, we further disclosed our findings through the bug bounty program. We made sure to keep the entire project confidential until we got the written approval to publish our results. 

We believe that the Ethereum community can benefit from the insights we are able to gain from the results of our deanonymization. Many aspects, such as geographical decentralization, could not be studied before. This improved understanding can further strengthen the security of the Ethereum blockchain.

Finally, while the lack of anonymity was noted before, researchers and engineers can benefit from the improved perspective we offer and can work on building a more robust and private gossiping mechanism for the future.

\Paragraph{Justice}
The benefits (and risks) of research should be fairly distributed. We did not single out specific individuals or organizations, and our work was conducted with the intent of benefiting the broader Ethereum community. To this end, we run \textsc{Rainbow} nodes across multiple locations. 

\Paragraph{Respect for Law and Public Interest}
In line with the principle of respect for law and public interest, we conducted our research in compliance with legal and regulatory frameworks surrounding blockchain technology and data privacy. No actions were taken to exploit the vulnerabilities found; instead, our research was conducted with the aim of improving the Ethereum network's security and privacy features.

\section{Open Science}
\label{sec:openscience}

As discussed in \Cref{sec:ethics}, we believe that our collected data is sensitive and should not be shared. Therefore, we only present aggregated data and refrain from any analysis targeting single entities. We also suggest keeping our \textsc{Rainbow} client private, until appropriate countermeasures have been implemented. Therefore, we refrain from making them publicly available. %

\section*{Acknowledgments}
We thank the anonymous USENIX Security 2025 reviewers for their useful suggestions. L. Kiffer contributed to this project while partially supported by the armasuisse Science and Technology CYD Distinguished Postdoctoral Fellowship. We also thank Amazon Web Services for a gift of AWS credit used towards the collection of our data.

\bibliographystyle{plain}
\bibliography{bibliography}

\begin{thebibliography}{10}

\bibitem{client2024prysm}
Ether Alpha.
\newblock {Client Diversity}.
\newblock \url{https://clientdiversity.org/#distribution}.
\newblock Accessed: 2024-08-10.

\bibitem{apostolaki2018sabre}
Maria Apostolaki, Gian Marti, Jan Müller, and Laurent Vanbever.
\newblock {SABRE: Protecting Bitcoin against Routing Attacks}.
\newblock In {\em Proceedings of the 26th Annual Network and Distributed System Security Symposium}, 2019.

\bibitem{apostolaki2017hijacking}
Maria Apostolaki, Aviv Zohar, and Laurent Vanbever.
\newblock {Hijacking Bitcoin: Routing Attacks on Cryptocurrencies}.
\newblock In {\em {IEEE Symposium on Security and Privacy (SP)}}, 2017.

\bibitem{ethereum_node_requirements_2024}
Bacloud.
\newblock {Ethereum Node Server Requirements 2024}.
\newblock \url{https://www.bacloud.com/en/knowledgebase/203/ethereum-node-server-requirements-2024.html}, 2024.
\newblock Accessed: 2025-01-01.

\bibitem{bailey2012menlo}
Michael Bailey, David Dittrich, Erin Kenneally, and Doug Maughan.
\newblock {The Menlo Report}.
\newblock {\em IEEE Symposium on Security and Privacy (SP)}, 2012.

\bibitem{basyoni2020survey}
Lamiaa Basyoni, Noora Fetais, Aiman Erbad, Amr Mohamed, and Mohsen Guizani.
\newblock {Traffic Analysis Attacks on Tor: A Survey}.
\newblock In {\em IEEE International Conference on Informatics, IoT, and Enabling Technologies (ICIoT)}, 2020.

\bibitem{beacon}
beaconcha.in.
\newblock \url{https://beaconcha.in/}, 2024.

\bibitem{biryukov2014deanonymisation}
Alex Biryukov, Dmitry Khovratovich, and Ivan Pustogarov.
\newblock {Deanonymisation of Clients in Bitcoin P2P Network}.
\newblock In {\em ACM SIGSAC Conference on Computer and Communications Security}, 2014.

\bibitem{biryukov2019deanonymization}
Alex Biryukov and Sergei Tikhomirov.
\newblock {Deanonymization and Linkability of Cryptocurrency Transactions Based on Network Analysis}.
\newblock In {\em IEEE European Symposium on Security and Privacy (EuroSP)}, 2019.

\bibitem{ethresearch2022dandelion}
Blagoj.
\newblock {Ethereum Consensus Layer Validator Anonymity Using Dandelion++ and RLN Conclusion}.
\newblock \url{https://ethresear.ch/t/ethereum-consensus-layer-validator-anonymity-using-dandelion-and-rln-conclusion/12698}, 2022.
\newblock Accessed: 2024-06-01.

\bibitem{bloxroute}
Bloxroute.
\newblock {Unlock Higher APYs and Increased Effectiveness with the Validator Gateway, bloXroute’s Performance Solution for Ethereum Validators \& Node Operators}.
\newblock \url{https://bloxroute.com/products/}.
\newblock Accessed: 2024-06-01.

\bibitem{bojja2017dandelion}
Shaileshh Bojja~Venkatakrishnan, Giulia Fanti, and Pramod Viswanath.
\newblock {Dandelion: Redesigning the Bitcoin Network for Anonymity}.
\newblock {\em Proceedings of the ACM on Measurement and Analysis of Computing Systems}, 2017.

\bibitem{boneh2020single}
Dan Boneh, Saba Eskandarian, Lucjan Hanzlik, and Nicola Greco.
\newblock {Single Secret Leader Election}.
\newblock In {\em 2nd ACM Conference on Advances in Financial Technologies}, 2020.

\bibitem{brown2024exploring}
Simon Brown and Leonardo Bautista-Gomez.
\newblock {Exploring Correlation Patterns in the Ethereum Validator Network}.
\newblock {\em arXiv preprint arXiv:2404.02164}, 2024.

\bibitem{buterin2017casper}
Vitalik Buterin and Virgil Griffith.
\newblock {Casper the Friendly Finality Gadget}.
\newblock {\em arXiv preprint arXiv:1710.09437}, 2017.

\bibitem{buterin2022secretelection}
Vitalike Buterin.
\newblock {Secret Non-Single Leader Election}.
\newblock \url{https://ethresear.ch/t/secret-non-single-leader-election/11789}, 2022.
\newblock Accessed: 2024-06-01.

\bibitem{hopr}
Sebastian Bürgel.
\newblock {Proof-of-Stake Validator Sniping Research}.
\newblock \url{https://medium.com/hoprnet/proof-of-stake-validator-sniping-research-8670c4a88a1c}, 2022.

\bibitem{cao2020exploring}
Tong Cao, Jiangshan Yu, J{\'e}r{\'e}mie Decouchant, Xiapu Luo, and Paulo Verissimo.
\newblock {Exploring the Monero Peer-to-Peer Network}.
\newblock In {\em Financial Cryptography and Data Security (FC)}, 2020.

\bibitem{DaianFlash2020}
Philip Daian, Steven Goldfeder, Tyler Kell, Yunqi Li, Xueyuan Zhao, Iddo Bentov, Lorenz Breidenbach, and Ari Juels.
\newblock {Flash Boys 2.0: Frontrunning in Decentralized Exchanges, Miner Extractable Value, and Consensus Instability}.
\newblock In {\em 2020 IEEE Symposium on Security and Privacy (SP)}, 2020.

\bibitem{d2022goldfish}
Francesco D'Amato, Joachim Neu, Ertem~Nusret Tas, and David Tse.
\newblock {Goldfish: No More Attacks on Proof-of-Stake Ethereum}.
\newblock {\em Financial Cryptography and Data Security (FC)}, 2024.

\bibitem{decker2013information}
Christian Decker and Roger Wattenhofer.
\newblock {Information Propagation in the Bitcoin Network}.
\newblock In {\em IEEE P2P}, 2013.

\bibitem{delgado2019txprobe}
Sergi Delgado-Segura, Surya Bakshi, Cristina P{\'e}rez-Sol{\`a}, James Litton, Andrew Pachulski, Andrew Miller, and Bobby Bhattacharjee.
\newblock {Txprobe: Discovering Bitcoin’s Network Topology Using Orphan Transactions}.
\newblock In {\em Financial Cryptography and Data Security (FC)}, 2019.

\bibitem{dingledine2004tor}
Roger Dingledine, Nick Mathewson, Paul~F Syverson, et~al.
\newblock Tor: The second-generation onion router.
\newblock In {\em USENIX Security Symposium (USENIX Security)}, 2004.

\bibitem{dmarzb2013committess}
dmarz.
\newblock Committess.
\newblock \url{https://hackmd.io/@dmarz/ethereum_overlays#Committess}.
\newblock Accessed: 2024-08-10.

\bibitem{dmarzb2013gossip}
dmarz.
\newblock {GossipSub Topics}.
\newblock \url{https://hackmd.io/@dmarz/ethereum_overlays#GossipSub-Topics}.
\newblock Accessed: 2024-08-10.

\bibitem{ethereum2021bls}
{Ethereum Foundation}.
\newblock Bls signatures.
\newblock \url{https://github.com/ethereum/consensus-specs/blob/dev/specs/phase0/bl-chain.md#bls-signatures }, 2024.
\newblock Accessed: 2024-06-03.

\bibitem{ethereum2024danebspecs}
{Ethereum Foundation}.
\newblock Deneb validator specifications.
\newblock \url{https://github.com/ethereum/consensus-specs/blob/dev/specs/deneb/validator.md}, 2024.
\newblock Accessed: 2024-06-03.

\bibitem{ethereum2024consensusspecs}
{Ethereum Foundation}.
\newblock Proof-of-stake (pos).
\newblock \url{https://ethereum.org/en/developers/docs/consensus-mechanisms/pos/}, 2024.
\newblock Accessed: 2024-06-03.

\bibitem{fanti2018dandelion++}
Giulia Fanti, Shaileshh~Bojja Venkatakrishnan, Surya Bakshi, Bradley Denby, Shruti Bhargava, Andrew Miller, and Pramod Viswanath.
\newblock {Dandelion++ Lightweight Cryptocurrency Networking with Formal Anonymity Guarantees}.
\newblock {\em Proceedings of the ACM on Measurement and Analysis of Computing Systems}, 2018.

\bibitem{fanti2017deanonymization}
Giulia Fanti and Pramod Viswanath.
\newblock {Deanonymization in the Bitcoin P2P Network}.
\newblock {\em Advances in Neural Information Processing Systems}, 2017.

\bibitem{figueiredo2004OnTA}
Daniel~R. Figueiredo, Philippe Nain, and Donald~F. Towsley.
\newblock {On the Analysis of the Predecessor Attack on Anonymity Systems}.
\newblock 2004.

\bibitem{ethereum2024dvt}
Ethereum Foundation.
\newblock {Distributed Validator Technology}.
\newblock \url{https://ethereum.org/en/staking/dvt/}.
\newblock Accessed: 2024-12-20.

\bibitem{discv5}
Ethereum Foundation.
\newblock {Node Discovery Protocol v5 - Theory}.
\newblock \url{https://github.com/ethereum/devp2p/blob/master/discv5/discv5-theory.md}.
\newblock Accessed: 2024-06-01.

\bibitem{eipsubnets}
Ethereum Foundation.
\newblock {Proposed Attnets Revamp \#2749}.
\newblock \url{https://github.com/ethereum/consensus-specs/issues/2749 }, 2021.

\bibitem{gao2019topology}
Yue Gao, Jinqiao Shi, Xuebin Wang, Qingfeng Tan, Can Zhao, and Zelin Yin.
\newblock {Topology Measurement and Analysis on Ethereum P2P Network}.
\newblock In {\em 2019 IEEE Symposium on Computers and Communications (ISCC)}, 2019.

\bibitem{gencer2018decentralization}
Adem~Efe Gencer, Soumya Basu, Ittay Eyal, Robbert Van~Renesse, and Emin~G{\"u}n Sirer.
\newblock {Decentralization in Bitcoin and Ethereum Metworks}.
\newblock In {\em Financial Cryptography and Data Security (FC)}, 2018.

\bibitem{giuliari2024dos}
Giacomo Giuliari, Alberto Sonnino, Marc Frei, Fabio Streun, Lefteris Kokoris-Kogias, and Adrian Perrig.
\newblock {An Empirical Study of Consensus Protocols’ DoS Resilience}, 2024.

\bibitem{grandjean2024ethereum}
Dominic Grandjean, Lioba Heimbach, and Roger Wattenhofer.
\newblock {Ethereum Proof-of-Stake Consensus Layer: Participation and Decentralization}.
\newblock In {\em 5th Workshop on Coordination of Decentralized Finance}, 2024.

\bibitem{guerraoui2023inherent}
Rachid Guerraoui, Anne-Marie Kermarrec, Anastasiia Kucherenko, Rafael Pinot, and Sasha Voitovych.
\newblock {On the Inherent Anonymity of Gossiping}.
\newblock In {\em 37th International Symposium on Distributed Computing (DISC)}, 2023.

\bibitem{heilman2015eclipse}
Ethan Heilman, Alison Kendler, Aviv Zohar, and Sharon Goldberg.
\newblock Eclipse attacks on bitcoin’s peer-to-peer network.
\newblock In {\em 24th USENIX Security Symposium (USENIX Security)}, 2015.

\bibitem{HeimbachEthereum2023}
Lioba Heimbach, Lucianna Kiffer, Christof Ferreira~Torres, and Roger Wattenhofer.
\newblock {Ethereum's Proposer-Builder Separation: Promises and Realities}.
\newblock In {\em ACM Internet Measurement Conference (IMC)}, 2023.

\bibitem{henningsen2019eclipsing}
Sebastian Henningsen, Daniel Teunis, Martin Florian, and Bj{\"o}rn Scheuermann.
\newblock {Eclipsing Ethereum Peers with False Friends}.
\newblock {\em arXiv preprint arXiv:1908.10141}, 2019.

\bibitem{multi_node_explorer}
Xin Hong and Lucianna Kiffer.
\newblock {Multi-Node Explorer}.
\newblock \url{http://multi-node-explorer.ethz.ch/cons}, 2025.

\bibitem{github2018brainstorming}
Jannikluhn.
\newblock {Brainstorming about Validator Privacy}.
\newblock \url{https://github.com/ethresearch/p2p/issues/5}.
\newblock Accessed: 2024-08-10.

\bibitem{ethresearch2022whisk}
George Kadianakis.
\newblock {Whisk: A Practical Shuffle-Based SSLE Protocol for Ethereum}.
\newblock \url{https://ethresear.ch/t/whisk-a-practical-shuffle-based-ssle-protocol-for-ethereum/11763}, 2022.
\newblock Accessed: 2024-06-01.

\bibitem{das2022tweet}
Sreeram Kannan.
\newblock Tweet.
\newblock \url{https://x.com/sreeramkannan/status/1563615609925304320?s=43&t=eDpUtjHy68wCYy40doq9RQ}, 2022.

\bibitem{kiffer2021under}
Lucianna Kiffer, Asad Salman, Dave Levin, Alan Mislove, and Cristina Nita-Rotaru.
\newblock {Under the Hood of the Ethereum Gossip Protocol}.
\newblock In {\em Financial Cryptography and Data Security (FC)}, 2021.

\bibitem{2022reorg}
Georgios Konstantopoulos and Vitalik Buterin.
\newblock {Ethereum Reorgs After The Merge}.
\newblock \url{https://www.paradigm.xyz/2021/07/ethereum-reorgs-after-the-merge}, 2021.

\bibitem{kumar2023verification}
Ankit Kumar, Max von Hippel, Panagiotis Manolios, and Cristina Nita-Rotaru.
\newblock {Verification of GossipSub in ACL2s}.
\newblock {\em arXiv preprint arXiv:2311.08859}, 2023.

\bibitem{kumar2023formal}
Ankit Kumar, Max von Hippel, Panagiotis Manolios, and Cristina Nita-Rotaru.
\newblock {Formal Model-Driven Analysis of Resilience of GossipSub to Attacks from Misbehaving Peers}.
\newblock In {\em IEEE Symposium on Security and Privacy (SP)}, 2024.

\bibitem{li2021toposhot}
Kai Li, Yuzhe Tang, Jiaqi Chen, Yibo Wang, and Xianghong Liu.
\newblock {TopoShot: Uncovering Ethereum's Network Topology Leveraging Replacement Transactions}.
\newblock In {\em Proceedings of the 21st ACM Internet Measurement Conference}, 2021.

\bibitem{libp2p2020dosmitigation}
Libp2p.
\newblock {DoS Mitigation}.
\newblock \url{https://docs.libp2p.io/concepts/security/dos-mitigation/}.
\newblock Accessed: 2024-08-10.

\bibitem{gossipsubspecs}
libp2p.
\newblock gossipsub v1.1.
\newblock \url{https://github.com/libp2p/specs/blob/master/pubsub/gossipsub/gossipsub-v1.1.md}, 2021.
\newblock Accessed: 2025-01-01.

\bibitem{lin2021measuring}
Qinwei Lin, Chao Li, Xifeng Zhao, and Xianhai Chen.
\newblock {Measuring Decentralization in Bitcoin and Ethereum Using Multiple Metrics and Granularities}.
\newblock In {\em IEEE 37th International Conference on Data Engineering Workshops (ICDEW)}, 2021.

\bibitem{marcus2018low}
Yuval Marcus, Ethan Heilman, and Sharon Goldberg.
\newblock {Low-Resource Eclipse Attacks on Ethereum's Peer-to-Peer Network}.
\newblock {\em Cryptology ePrint Archive}, 2018.

\bibitem{maymounkov2002kademlia}
Petar Maymounkov and David Mazieres.
\newblock {Kademlia: A Peer-to-Peer Information System Based on the XOR Metric}.
\newblock In {\em International Workshop on Peer-to-Peer Systems}, 2002.

\bibitem{vouch2024}
Jim McDonald.
\newblock {Vouch Validator Client}.
\newblock \url{https://www.attestant.io/posts/introducing-vouch/}.
\newblock Accessed: 2024-06-01.

\bibitem{miller2015discovering}
Andrew Miller, James Litton, Andrew Pachulski, Neal Gupta, Dave Levin, Neil Spring, Bobby Bhattacharjee, et~al.
\newblock {Discovering Bitcoin’s Public Topology and Influential Nodes}.
\newblock 2015.

\bibitem{murdoch2005low}
Steven~J Murdoch and George Danezis.
\newblock {Low-Cost Traffic Analysis of Tor}.
\newblock In {\em IEEE Symposium on Security and Privacy (SP)}, 2005.

\bibitem{nasr2018deepcorr}
Milad Nasr, Alireza Bahramali, and Amir Houmansadr.
\newblock {Deepcorr: Strong Flow Correlation Attacks on Tor Using Deep Learning}.
\newblock In {\em ACM SIGSAC Conference on Computer and Communications Security}, 2018.

\bibitem{neudecker2016timing}
Till Neudecker, Philipp Andelfinger, and Hannes Hartenstein.
\newblock {Timing Analysis for Inferring the Topology of the Bitcoin Peer-to-Peer Network}.
\newblock In {\em 2016 Intl IEEE Conferences on Ubiquitous Intelligence \& Computing, Advanced and Trusted Computing, Scalable Computing and Communications, Cloud and Big Data Computing, Internet of People, and Smart World Congress (UIC/ATC/ScalCom/CBDCom/IoP/SmartWorld)}. IEEE, 2016.

\bibitem{eip7251}
Mike Neuder, Francesco d’Amato, Aditya Asgaonkar, and Justin Drake.
\newblock {EIP-7251: Increase MAX\_EFFECTIVE\_BALANCE to 2048 ETH}.
\newblock \url{https://github.com/ethereum/EIPs/blob/master/EIPS/eip-7251.md}, 2024.
\newblock Accessed: 2024-09-05.

\bibitem{das2023a16z}
Valeria Nikolaenko and Dan Boneh.
\newblock {Data Availability Sampling and Danksharding: An Overview and a Proposal for Improvements}.
\newblock \url{https://a16zcrypto.com/posts/article/an-overview-of-danksharding-and-a-proposal-for-improvement-of-das/}, 2023.

\bibitem{panchenko2007crowds}
Andriy Panchenko and Lexi Pimenidis.
\newblock {Crowds Revisited: Practically Effective Predecessor Attack}.
\newblock In {\em 12th Nordic Workshop on Secure IT-Systems (NordSec 2007)}, 2007.

\bibitem{bankless2024pectra}
William~M. Peaster.
\newblock {Ethereum’s Pectra EIPs: What You Need to Know}.
\newblock \url{https://www.bankless.com/ethereum-pectra-eips}, 2024.

\bibitem{prysmaticlabs2024prysm}
Prysmaticlabs.
\newblock Prysm.
\newblock \url{https://github.com/prysmaticlabs/prysm}.
\newblock Accessed: 2024-08-10.

\bibitem{QinQuantifying2022}
Kaihua Qin, Liyi Zhou, and Arthur Gervais.
\newblock {Quantifying Blockchain Extractable Value: How Dark is the Forest?}
\newblock In {\em IEEE Symposium on Security and Privacy (SP)}, 2022.

\bibitem{ethresearch2020packetology}
TXRX Research.
\newblock {Packetology: Validator Privacy}.
\newblock \url{https://ethresear.ch/t/packetology-validator-privacy/7547}.
\newblock Accessed: 2024-08-10.

\bibitem{rohrer2020counting}
Elias Rohrer and Florian Tschorsch.
\newblock {Counting Down Thunder: Timing Attacks on Privacy in Payment Channel Networks}.
\newblock In {\em Proceedings of the 2nd ACM Conference on Advances in Financial Technologies}, 2020.

\bibitem{andras2023tor}
István~András Seres.
\newblock {Quantifying the Privacy Guarantees of Validatory Privacy Mechanisms}.
\newblock \url{https://ethresear.ch/t/quantifying-the-privacy-guarantees-of-validatory-privacy-mechanisms/15715}, 2023.
\newblock Accessed: 2024-06-01.

\bibitem{shmatikov2006timing}
Vitaly Shmatikov and Ming-Hsiu Wang.
\newblock {Timing Analysis in Low-Latency Mix Networks: Attacks and Sefenses}.
\newblock In {\em 11th European Symposium on Research in Computer Security (ESORICS)}, 2006.

\bibitem{tang2023strategic}
Weizhao Tang, Lucianna Kiffer, Giulia Fanti, and Ari Juels.
\newblock {Strategic Latency Reduction in Blockchain Peer-to-Peer Networks}.
\newblock {\em Proceedings of the ACM on Measurement and Analysis of Computing Systems}, 2023.

\bibitem{tran2020stealthier}
Muoi Tran, Inho Choi, Gi~Jun Moon, Anh~V Vu, and Min~Suk Kang.
\newblock {A Stealthier Partitioning Attack Against Bitcoin Peer-to-Peer Network}.
\newblock In {\em IEEE Symposium on Security and Privacy (SP)}, 2020.

\bibitem{tran2021routing}
Muoi Tran, Akshaye Shenoi, and Min~Suk Kang.
\newblock {On the Routing-Aware Peering Against Network-Eclipse Attacks in Bitcoin}.
\newblock In {\em 30th USENIX Security Symposium (USENIX Security)}, 2021.

\bibitem{tran2024routing}
Muoi Tran, Theo von Arx, and Laurent Vanbever.
\newblock {Routing Attacks on Cryptocurrency Mining Pools}.
\newblock In {\em 45th IEEE Symposium on Security and Privacy (SP)}, 2024.

\bibitem{vyzovitis2020gossipsub}
Dimitris Vyzovitis, Yusef Napora, Dirk McCormick, David Dias, and Yiannis Psaras.
\newblock {Gossipsub: Attack-Resilient Message Propagation in the Filecoin and ETH2.0 Networks}.
\newblock {\em arXiv preprint arXiv:2007.02754}, 2020.

\bibitem{vyzovitis2022gossipsub}
Dimitris Vyzovitis, Yusef Napora, Dirk McCormick, David Dias, and Yiannis Psaras.
\newblock {Gossipsub-v1. 1 Evaluation Report}, 2022.

\bibitem{mevboost_pics}
Toni Wahrstätter.
\newblock {MEV Boost Dashboard}.
\newblock \url{https://mevboost.pics/}, 2025.
\newblock Accessed: 2025-01-01.

\bibitem{wright2004predecessor}
Matthew~K. Wright, Micah Adler, Brian~Neil Levine, and Clay Shields.
\newblock {The Predecessor Attack: An Analysis of a Threat to Anonymous Communications Systems}.
\newblock {\em ACM Trans. Inf. Syst. Secur.}, 2004.

\end{thebibliography}

\clearpage
\appendix

\section{Heuristics} \label{app:heuristics}

To test the robustness of our heuristics, in Tables~\ref{tab:validators_heuristics} and~\ref{tab:label_consistency_heuristics} we investigate the impact of each condition on the effectiveness and accuracy of our deanonymization technique. 

Table~\ref{tab:validators_heuristics} analyzes how many validators we locate in total, as well as excluding those associated with service providers in columns one and two. We further use the verification method introduced in \Cref{sec:unique} to indicate how many non-unique validators we find and what proportion this makes up of the validators (excl. service providers). Each of the conditions is tested separately in the respective subtables.

In Table~\ref{tab:label_consistency_heuristics}, we extend the robustness testing of our parameter choices by employing the verification metrics introduced in \Cref{sec:service}. We measure the proportion of our peers who are assigned a validator set that is consistent, inconsistent, or unknown (validators whose labels do not pass consistency checks). We also present the same statistics for validators as opposed to peers, e.g., how many validators are part of a consistent validator set. Again, each of the conditions is tested separately in the respective subtables.

\begin{table*}[t]
\begin{subtable}[t]{\textwidth}
\centering
 \resizebox{0.95\columnwidth}{!}{
\begin{tabular}{cccc|rrrr}
\toprule
C1 & C2 & C3 & C4 & validators & validators (excl. service providers) & non-unique validators & \% non-unique validators \\
\midrule
\textbackslash & 1 & 10 & 1 & 355,493 & 271,260 & 35,676 & 13.15 \\
0.30 & 1 & 10 & 1 & 262,408 & 165,083 & 16,487 & 9.99 \\
\textbf{0.90} & \textbf{1} & \textbf{10} & \textbf{1} & 252,895 & 161,057 & 16,172 & 10.04 \\
1.00 & 1 & 10 & 1 & 208,481 & 125,578 & 7,150 & 5.69 \\
\bottomrule
\end{tabular}
}\caption{C1}
\label{tab:valc1}
\end{subtable}\vspace{2pt}

\begin{subtable}[t]{\textwidth}
\centering
 \resizebox{0.95\columnwidth}{!}{
\begin{tabular}{cccc|rrrr}
\toprule
C1 & C2 & C3 & C4 & validators & validators (excl. service providers) & non-unique validators & \% non-unique validators \\
\midrule
0.90 & \textbackslash  & 10 & 1 & 270,242 & 181,494 & 20,117 & 11.08 \\
\textbf{0.90} & \textbf{1} & \textbf{10} & \textbf{1} & 252,895 & 161,057 & 16,172 & 10.04 \\
\bottomrule
\end{tabular}
}\caption{C2}
\label{tab:valc2}
\end{subtable}\vspace{2pt}

\begin{subtable}[t]{\textwidth}
\centering
 \resizebox{0.95\columnwidth}{!}{
\begin{tabular}{cccc|rrrr}
\toprule
C1 & C2 & C3 & C4 & validators & validators (excl. service providers) & non-unique validators & \% non-unique validators \\
\midrule
0.90 & 1 & \textbackslash & 1 & 893,509 & 885,864 & 473,193 & 53.42 \\
0.90 & 1 & 20 & 1 & 376,088 & 270,441 & 29,111 & 10.76 \\
\textbf{0.90} & \textbf{1} & \textbf{10} & \textbf{1} & 252,895 & 161,057 & 16,172 & 10.04 \\
0.90 & 1 & 5 & 1 & 150,219 & 110,896 & 11,906 & 10.74 \\
\bottomrule
\end{tabular}
}\caption{C3}
\label{tab:valc3}
\end{subtable}\vspace{2pt}

\begin{subtable}[t]{\textwidth}
\centering
 \resizebox{0.95\columnwidth}{!}{
\begin{tabular}{cccc|rrrr}
\toprule
C1 & C2 & C3 & C4 & validators & validators (excl. service providers) & non-unique validators & \% non-unique validators \\
\midrule
0.90 & 1 & 10 & \textbackslash & 256,725 & 162,168 & 16,350 & 10.08 \\
\textbf{0.90} & \textbf{1} & \textbf{10} & \textbf{1} & 252,895 & 161,057 & 16,172 & 10.04 \\
0.90 & 1 & 10 & 4 & 159,494 & 136,350 & 15,308 & 11.23 \\
\bottomrule
\end{tabular}
}\caption{C4}
\label{tab:valc4}
\end{subtable}
\caption{Number of validators located by each \textsc{Rainbow} node and overall. The first column indicates the total number of validators, the second column excludes validators from peers we identify to be P2P service providers (see Section~\ref{sec:service}), and the third column indicates the validators with a non-unique mapping to an IP port combination (see Section~\ref{sec:unique}). We repeat the analysis for various parameter configurations of our heuristics, testing the parameter for each condition individually. Our chosen configuration is highlighted in bold.}\label{tab:validators_heuristics}\vspace{-6pt}
\end{table*}

\subsection{Condition C1}

Condition C1 ensures that the proportion of non-backbone attestations remains sufficiently large. The parameter $c1$ dictates how closely the observed ratio must align with the expected ratio. For example, when $c1$ is set to 0.9, the observed proportion must reach at least 90\% of the expected value to satisfy the condition.

Our analysis compares scenarios without condition C1 to those with three different $c1$ values (0.3, 0.9, and 1.0). We observe that omitting condition C1 results in a nearly one-third increase in the total number of validators and an almost 50\% increase when excluding service providers (see Table~\ref{tab:valc1}). Alarmingly, the number of non-unique validators more than doubles, leading to a higher overall proportion of non-unique validators. Turning to Table~\ref{tab:c1}, we find that without condition C1, the consistency of validator sets diminishes significantly, further highlighting the importance of this condition as part of our deanonymization methodology.

For the remaining values we test for C1, we observe smaller differences. Even with $c1$ set to 0.3, we observe similar numbers across the board as with $c1$ set to 0.9 (the parameter we choose). In general, we find slightly fewer validators with $c1$ set to 0.9, but observe that the validator sets we locate are more consistent. For the even more restrictive configuration ($c1=1.0$), we see that we lose a significant number of validators and have a smaller proportion of non-unique validators (see Table~\ref{tab:valc1}). At the same time, the validator sets do not grow more consistent (see Table~\ref{tab:c1}).

In conclusion, our analysis demonstrates that condition C1 is essential for reducing likely false positives, as evidenced by the significant increase in non-unique validators and inconsistency when C1 is absent. We select $c1=0.9$ as it strikes a balance between being restrictive and accommodating minor deviations from the expected ratio, allowing for small discrepancies without compromising the overall consistency.

\subsection{Condition C2}

Condition C2 ensures that the node is not subscribed to all subnets. If a node were subscribed to all subnets, it would be impossible to receive non-backbone attestations from that node, rendering condition C1 ineffective.

We perform the analysis both without and with (indicated by a 1 in Table~\ref{tab:valc1} and Table~\ref{tab:c2}) condition C2. Without C2, we observe a slight rise in the number of validators, but at a small cost in the proportion of non-unique validators (see Table~\ref{tab:valc2}) and a slight increase in the number of inconsistent validator sets on peers (see Table~\ref{tab:c2}). 
Thus, we use condition C2 to avoid having to rely on less obvious signals (i.e., number of messages as opposed to type of message. 

\begin{table*}[th]
\begin{subtable}[t]{\textwidth}
\centering
 \resizebox{0.8\columnwidth}{!}{
\begin{tabular}{rrrr|rrr|rrr}
\toprule
C1 & C2 & C3 & C4 & \multicolumn{3}{c}{peers} & \multicolumn{3}{c}{validators} \\
 &  &  &  & consistent & inconsistent & unknown & consistent & inconsistent & unknown \\
\midrule
\textbackslash & 1 & 10 & 1 & 74.32\% & 19.92\% & 5.75\% & 47.15\% & 55.73\% & 1.50\% \\
0.30 & 1 & 10 & 1 & 92.56\% & 1.65\% & 5.79\% & 61.84\% & 41.50\% & 2.10\% \\
\textbf{0.90} & \textbf{1} & \textbf{10} & \textbf{1} & 93.75\% & 0.92\% & 5.33\% & 63.67\% & 39.67\% & 2.24\% \\
1.00 & 1 & 10 & 1 & 93.38\% & 0.77\% & 5.85\% & 58.18\% & 44.24\% & 2.93\% \\
\bottomrule
\end{tabular}
}\caption{C1}
\label{tab:c1}
\end{subtable}

\begin{subtable}[t]{\textwidth}
\centering
 \resizebox{0.8\columnwidth}{!}{
\begin{tabular}{rrrr|rrr|rrr}
\toprule
C1 & C2 & C3 & C4 & \multicolumn{3}{c}{peers} & \multicolumn{3}{c}{validators} \\
 &  &  &  & consistent & inconsistent & unknown & consistent & inconsistent & unknown \\
\midrule
0.90 & \textbackslash & 10 & 1 & 93.73\% & 0.97\% & 5.30\% & 65.21\% & 37.98\% & 2.11\% \\
\textbf{0.90} & \textbf{1} & \textbf{10} & \textbf{1} & 93.75\% & 0.92\% & 5.33\% & 63.67\% & 39.67\% & 2.24\% \\
\bottomrule

\end{tabular}
}\caption{C2}
\label{tab:c2}
\end{subtable}

\begin{subtable}[t]{\textwidth}
\centering
 \resizebox{0.8\columnwidth}{!}{
\begin{tabular}{rrrr|rrr|rrr}
\toprule
C1 & C2 & C3 & C4 & \multicolumn{3}{c}{peers} & \multicolumn{3}{c}{validators} \\
 &  &  &  & consistent & inconsistent & unknown & consistent & inconsistent & unknown \\
\midrule
0.90 & 1 & \textbackslash & 1 & 66.17\% & 27.70\% & 6.13\% & 7.70\% & 94.71\% & 0.23\% \\
0.90 & 1 & 20 & 1 & 88.37\% & 6.29\% & 5.33\% & 34.88\% & 66.86\% & 1.01\% \\
\textbf{0.90} & \textbf{1} & \textbf{10} & \textbf{1} & 93.75\% & 0.92\% & 5.33\% & 63.67\% & 39.67\% & 2.24\% \\
0.90 & 1 & 5 & 1 & 94.03\% & 0.65\% & 5.32\% & 78.19\% & 23.25\% & 3.20\% \\

\bottomrule

\end{tabular}
}\caption{C3}
\label{tab:c3}
\end{subtable}

\begin{subtable}[t]{\textwidth}
\centering
 \resizebox{0.8\columnwidth}{!}{
\begin{tabular}{rrrr|rrr|rrr}
\toprule
C1 & C2 & C3 & C4 & \multicolumn{3}{c}{peers} & \multicolumn{3}{c}{validators} \\
 &  &  &  & consistent & inconsistent & unknown & consistent & inconsistent & unknown \\
\midrule
0.90 & 1 & 10 & \textbackslash & 93.76\% & 0.92\% & 5.33\% & 63.21\% & 40.13\% & 2.24\% \\
\textbf{0.90} & \textbf{1} & \textbf{10} & \textbf{1} & 93.75\% & 0.92\% & 5.33\% & 63.67\% & 39.67\% & 2.24\% \\
0.90 & 1 & 10 & 4 & 93.83\% & 0.83\% & 5.34\% & 86.28\% & 10.74\% & 3.65\% \\
\bottomrule

\end{tabular}
}\caption{C4}
\label{tab:c4}
\end{subtable}
\caption{Percentage of peers with a consistent validator set and validators belonging to a consistent validator set across our four \textsc{Rainbow} nodes as well as overall. Consistency is defined in \Cref{sec:service}. We repeat the analysis for various parameter configurations of our heuristics, testing the parameter for each condition individually. Our chosen configuration is highlighted in bold. }\label{tab:label_consistency_heuristics}\vspace{-6pt}
\end{table*}

\subsection{Condition C3}

With condition C3 we ensure that we get at least every $c3$-th message from a peer for a specific validator that we locate on the peer. This is important to ensure that we receive enough data and maintain a good connection with the peer. 

We test the condition C3 by repeating the analysis without C3 and by using three different parameter configurations (i.e., 20, 10, and 5). The analysis reveals that in the absence of C3 the number of validators is more than three-fold and the number of non-unique validators skyrockets (see Table~\ref{tab:valc3}). Further, the consistency of validator sets on peers decreases significantly (see Table~\ref{tab:c3}). This underscores the importance of having a good enough connection as a prerequisite for our deanonymization methodology. 

For the various parameters for $c3$ we test, the differences are less extreme but still significant. With $c3=20$, the least restrictive parameter combination we test, we increase the validator count by almost one-third without significant increases in terms of the proportion of non-unique validators, but with a noticeable decrease in the proportion of consistent peers in comparison to our chosen parameter combination ($c3=10$). When comparing $c3=10$ to $c3=5$, we observe that with $c3=10$ we find significantly more validators but experience almost no losses in terms of the proportion of non-unique validators or the proportion of peers with consistent validator sets. We only observe that the proportion of validators in consistent validator sets decreases. However, this is due to more peers being associated with P2P service providers than otherwise (see Section~\ref{sec:service}). As these are excluded in our analysis, there is no significant negative impact of choosing the slightly less restrictive parameter $c3=10$. 

To summarize, condition C3 ensures that we only try to deanonymize validators on a peer if we were able to maintain a good connection. Otherwise, false positives might become more likely, as indicated by the rise in non-unique validators and inconsistent validator sets in the absence of condition C3.

\subsection{Condition C4}

Our final condition C4 ensures that we receive significantly more messages from a validator we map to a peer than for the remaining validators for which we receive messages from the same peer. In particular, we require that the number of messages received for a validator we deanonymize on a peer is $c4$ standard deviations more than for the mean across all validators for which we receive messages from the peer. This heuristic is an extra precautious measure to ensure that there is a separation in terms of the number of messages between validators mapped to the peer and those not. 

To test this condition, we perform the analysis without C4 and for two values for $c4$, namely 1 and 4. We find that the results appear largely similar for all configurations in terms of the proportion of non-unique validators (see Table~\ref{tab:valc4}) and consistency (see Table~\ref{tab:c4}). The main difference is that for the most restrictive $c4$ value of 4, we locate significantly fewer validators. Thus, we choose $c4=1$ for our parameter configuration. 

Note that even though this heuristic has the smallest impact, as revealed by our testing, we have the heuristic in place as a measure of caution.

\subsection{Fanout Membership and False-Negatives}\label{app:fanout}

To deanonymize a peer, we rely on receiving attestations that originate from that peer (i.e., attestations signed by a validator running on that peer's machine) on subnets where the peer is currently not participating as a backbone. To receive these non-backbone attestations, the peer must have added us to their \textit{fanout} for that subnet. In our deanonymization approach, we consider only peers for which we are included in several fanouts (controlled by Condition C3). We note that if we ever participate in \textit{any} fanout -- i.e., by receiving at least \textit{one} non-backbone attestation\footnote{This could also happen during dynamic subset membership, which implies a validator that is aggregating.}-- this peer is considered as potentially hosting a validator, and is thus either deanonymized, or included in the "rest" column of Table~\ref{tab:deanon} (those peers that would need more careful examination). We, therefore, only consider peers who have \textit{never} sent a non-backbone attestation (and are not in all 64 subnets) as running no validators. 

We now argue the unlikelihood of never receiving a non-backbone attestation from a peer running a validator based on the gossip protocol specs~\cite{gossipsubspecs}.\footnote{We use the default parameters of the Prysm client but note other implementations have similar parameters.} When we first connect to the peer, they add us to all fanouts that are not at capacity (default fanout size is 8) that we are a backbone in (i.e., all of them). Thus, a peer participating in 2 subnets will have 62 fanouts. Assuming they are running with a default peer cap of 50-100 peers all on 2 subnets, in expectation each fanout would have less than 4 peers (be half empty), thus we would be added to all of them. Even allowing for some of their peers running on all 64 subnets,\footnote{Note that less than 1\% of our peers run on all subnets (see Table~\ref{tab:deanon}).} this would still leave most fanouts not at capacity in expectation. Thus, when we connect for the first time to a peer, we are quite likely added to at least one fanout. A peer running with non-default parameters may run with higher peer counts and thus have full fanouts. These could lead to false negatives in our approach. We note that the longer we maintain a connection to a peer, the stronger the likelihood of eventually being added to one fanout long enough to receive at least one non-backbone attestation.

\end{document}